\documentclass[aps,prx,twocolumn,10pt,amssymb,amsmath,superscriptaddress,longbibliography]{revtex4-2} 
\usepackage{amsfonts,amsmath,amssymb,graphicx}
\usepackage{cancel}
\usepackage{color,times}
\definecolor{darkblue}{rgb}{0, 0, 0.8}
\usepackage{grffile}
\usepackage{braket}
\usepackage{bm}

\usepackage[colorlinks=true, breaklinks=true, linkcolor=red, citecolor=darkblue, urlcolor=darkblue]{hyperref}

\newcommand{\figref}[1]{Fig.~\ref{#1}}
\renewcommand{\eqref}[1]{Eq.~(\ref{#1})}

\newcommand{\E}{\mathrm{e}}
\newcommand{\I}{\mathrm{i}}
\newcommand{\dd}{{\rm d}}
\newcommand{\pdag}{\phantom{\dagger}}

\def\nnb{\langle i,j \rangle}
\def\2nb{\langle\langle i,j \rangle\rangle}
\def\beq{\begin{equation}}
\def\eeq{\end{equation}}
\def\bea{\begin{eqnarray}}
\def\eea{\end{eqnarray}}
\def\ba{\begin{array}{ccc}}
\def\ea{\end{array}}
\def\nn{\nonumber \\}
\def\ns{N_s}

\graphicspath{{figures/}}

\begin{document}

\title{Torus Spectroscopy of the Gross-Neveu-Yukawa Quantum Field Theory:\\ Free Dirac versus Chiral Ising Fixed Point }

\author{Michael Schuler}
\affiliation{Institut f\"ur Theoretische Physik, Universit\"at Innsbruck, A-6020 Innsbruck, Austria}
\affiliation{Vienna Center for Quantum Science and Technology, Atominstitut, TU Wien, 1040 Wien, Austria}
\author{Stephan Hesselmann}
\affiliation{Institut f\"ur Theoretische Festk\"orperphysik, JARA-FIT and JARA-HPC,
RWTH Aachen University, 52056 Aachen, Germany}
\author{Seth Whitsitt}
\affiliation{Joint Quantum Institute, National Institute of Standards and Technology and the University of Maryland, College Park, MD, 20742, USA}
\author{Thomas C. Lang}
\affiliation{Institut f\"ur Theoretische Physik, Universit\"at Innsbruck, A-6020 Innsbruck, Austria}
\author{Stefan Wessel}
\affiliation{Institut f\"ur Theoretische Festk\"orperphysik, JARA-FIT and JARA-HPC,
RWTH Aachen University, 52056 Aachen, Germany}
\author{Andreas M. L\"auchli}
\affiliation{Institut f\"ur Theoretische Physik, Universit\"at Innsbruck, A-6020 Innsbruck, Austria}

\date{\today}

\begin{abstract}
We establish the universal torus low-energy spectra at the free Dirac fixed point and at the strongly coupled {\em chiral Ising} fixed point and their subtle crossover behaviour in the Gross-Neuveu-Yukawa field theory with ${n_\text{D}=4}$ component Dirac spinors in $D=(2+1)$ dimensions. These fixed points and the field theories are directly relevant for the long-wavelength physics of certain interacting Dirac systems, such as repulsive spinless fermions on the honeycomb lattice or $\pi$-flux square lattice. The torus energy spectrum has been shown previously to serve as a characteristic fingerprint of relativistic fixed points and is a powerful tool to discriminate quantum critical behaviour in numerical simulations. Here, we use a combination of exact diagonalization and quantum Monte Carlo simulations of strongly interacting fermionic lattice models, to compute the critical torus energy spectrum on finite-size clusters with periodic boundaries and extrapolate them to the thermodynamic limit. Additionally, we compute the torus energy spectrum analytically using the perturbative expansion in ${\epsilon = 4 - D}$, which is in good agreement with the numerical results, thereby validating the presence of the chiral Ising fixed point in the lattice models at hand. We show that the strong interaction between the spinor field and the scalar order-parameter field strongly influences the critical torus energy spectrum and we observe prominent multiplicity features related to an emergent symmetry predicted from the quantum field theory. Building on these results we are able to address the subtle crossover physics of the low-energy spectrum flowing from the chiral Ising fixed point to the Dirac fixed point, and analyze earlier flawed attempts to extract Fermi velocity renormalizations from the low-energy spectrum.
\end{abstract}

\maketitle

\section{Introduction}
\label{sec:introduction}

Dirac fermions with a quasi-relativistic, i.e., gapless and linear, dispersion relation arise as low-energy quasi-particles in many condensed matter systems such as graphene and $d$-wave superconductors~\cite{Vafek2013,Herbut2006,Herbut2009}. Interactions between the fermions can drive the system from the Dirac semimetallic (SM)  phase through a quantum critical point (QCP) into various symmetry broken phases. Spinless fermions on the honeycomb or $\pi$-flux square lattice with repulsive nearest-neighbour interactions, for example, exhibit a SM to Mott insulator transition, where the ground state is charge ordered and spontaneously breaks  discrete sublattice exchange symmetries~\cite{Herbut2006,Li2015}.  The quantum critical point of such phase transitions involves fermionic degrees of freedom and is believed to be described by the chiral Ising fixed point of the $D=(2+1)$ dimensional Gross-Neveu-Yukawa (GNY) field theory which features strong coupling between fermionic spinors and a scalar field~\cite{zinn1991,Rosenstein1993,Herbut2009,Iliesiu2015,Ihrig2018}. Such universality classes do not have classical Landau-Ginzburg-Wilson analogues and are thus of particular interest.
Recently, many attempts have been made to precisely measure the scaling dimensions of the operators of chiral QCPs in GNY field theories~\cite{Li2015,Iliesiu2015,Knorr2016,Zerf2017,Iliesiu2018,xu2019quantum,Liu2020,Huffman2020,Tada2020}, which are a unique identifier of the universality class and directly related to the critical exponents of the phase transition. This task turned out to be particularly challenging and different methods, which were successful for charting more common critical points in the past, could not yet obtain completely consistent scaling dimensions or critical exponents. This summarizes the current situation for both the chiral Ising, and even more severely, the chiral Heisenberg universality class~\cite{Zerf2017, Knorr2018}.  

Here, we strike a new path to tackle this problem. In fact, another way to identify and chart universality classes is to measure their critical torus energy spectrum as it was shown in Refs.~\cite{Schuler2016, Whitsitt2016, Whitsitt2017,Thomson2017} for Wilson-Fisher and topological phase transitions. The low-energy gaps at a relativistic critical point on the torus are given, up to a non-universal factor $v$ describing the effective speed of light, by universal numbers $\xi_i$ times $1/L$,  where $L$ is the linear extent of the cluster \footnote{More generally, for a QCP with dynamic critical exponent $z$, the finite size spectrum takes the form $E_i = v \xi_i/L^z$, where $v$ is a non-universal pre-factor and the dimensionless numbers $\xi_i$ are universal constants. In this paper, we specialize to the relativistic case $z=1$, where $v$ is equal to the velocity of low-energy excitations}.  The order and degeneracy of the $\xi_i$ together with quantum numbers of the corresponding eigenstate (e.g.~momentum, fermion number) provide a unique fingerprint of the universality class and can be obtained by many complementary numerical and analytical techniques. 

Given the reported tension in the literature we want to shed light on the nature of the universality class from a different angle by confronting and comparing numerical torus energy spectra with analytical results. In this work, we use exact diagonalization (ED) and quantum Monte Carlo (QMC) simulations of fermionic lattice models, as well as $\epsilon$ expansion calculations of the effective low-energy field theory, to compute the critical torus energy spectrum for the chiral Ising universality class with an $(n_\text{D}=4)$-component spinor field. Although the $\epsilon$ expansion is only performed to low order, it provides important, exact statements about non-trivial multiplicities and quantum numbers of the low-energy spectrum in the scaling limit, while we provide high-quality data of the $\xi_i$ from numerical simulations.

Also, we want to emphasize that, while the $\xi_i$ are universal numbers, it is not only their precise values but typically the sequence of the low-energy levels with their degeneracies and quantum numbers which make the critical torus spectrum a universal fingerprint. In particular, changing the nature of the QCP by using another $n_D\neq4$ would lead to a different multiplicity structure, i.e. a qualitative change.
This is one of the main advantages of the critical torus spectrum compared to measuring critical exponents, where often very high precision data is necessary to distinguish different universality classes. 

Furthermore, the study of the GNY field theory and the corresponding microscopic models allows us to clarify the crossover flow of the torus energy spectrum between two different infrared (IR) fixed points. This advance enables us to reliably measure the Fermi velocity, which is the condensed matter analogy of the speed of light of the GNY field theory, a topic of recent controversy~\cite{Lang2019,Tang2018,Hesselmann2019}.

Our work also completes an important intermediate step towards a quantitative understanding of massless Dirac fermions coupled to a $U(1)$ gauge field (QED$_3$), which are of paramount importance for many quantum spin liquid candidates and exotic quantum phase transitions~\cite{Ran2007,Iqbal2013,He2017,Song2019,Senthil2004}.

The paper is organized as follows. In Sec.~\ref{sec:models} we introduce the GNY field theory, as well as the related Gross-Neveu (GN) field theory, which is formulated purely in terms of interacting fermions and introduce the two distinct renormalization group fixed points under consideration in this work. We present the fermion lattice models used to compute the chiral Ising critical torus spectrum, and establish the GNY field theory as a low-energy effective description of the lattice models.
In Sec.~\ref{sec:overview} we provide a brief overview of our results: We discuss the different structures of the torus spectra of the free Dirac conformal field theory (CFT) and chiral Ising CFT, the crossover behaviour in finite volume and their impact on the renormalization of the Fermi velocity. In Sec.~\ref{sec:hc_numerics} we give a more detailed analysis of the critical torus energy spectrum obtained from numerics. We show energy gaps from both ED and QMC simulations, and give details on the extrapolation to the thermodynamic limit. In Sec.~\ref{sec:eps_expansion} we present the $\epsilon$ expansion of the GNY field theory, and compare the results to the numerical spectra. Finally, in Sec.~\ref{sec:conclusion} we conclude our results by comparing the different torus geometries among each other and discuss possible future perspectives.

\section{Field Theories and Model Hamiltonians}
\label{sec:models}

This section provides a concise introduction of the GNY field theory, the important infrared fixed points along with symmetry aspects that are relevant for the analysis of  the torus spectrum. We also introduce the microscopic quantum many-body lattice models that exhibit fermionic quantum critical points,  and which we examine by our numerical methods.   

\subsection{Quantum Field Theories}
\label{sec:qft}

The fermionic quantum field theories (QFTs) that we explore in this work can be described by the GNY theory of fermionic fields coupled to a $\mathbb{Z}_2$ order parameter, i.e., a real, one-component scalar field~\cite{Herbut2006, Herbut2009b, Ihrig2018} in $D=3$ (space-time) dimensions. Depending on the value of a tuning parameter $s$, the GNY theory describes a SM of non-interacting Dirac fermions, a symmetry broken phase with finite order parameter, and, in between those, a critical point belonging to the \textit{chiral Ising} universality class~\cite{zinn1991,Rosenstein1993}. The most general form of the imaginary-time GNY Lagrangian is
\begin{equation}
    \mathcal{L}_\text{GNY} = - \overline{\Psi}^{j} \left( \cancel\partial + g_{\rm Y} \, \phi \right) \Psi^{j} + \frac{1}{2} \phi \left(
    s - \partial^2\right) \phi + \frac{\lambda}{4!} \phi^4\,,
    \label{eq:LGNY}
\end{equation}
where $\Psi^{j}$ is an $n_\text{D}$-component Dirac spinor with ${j = 1,...,{N_f}}$ flavors, so the total number of fermionic degrees of freedom is ${N = n_\text{D} N_f}$. The real scalar field is denoted by $\phi$, and $g_{\rm Y}$ is the Yukawa coupling strength between the spinor and scalar fields. We use the standard notation ${\cancel\partial = \gamma^\mu \partial_\mu}$, ${\mu \in \{0, \dots D-1\}}$ and ${\overline{\Psi} = \Psi \gamma^0}$, where the $\gamma^{\mu}$ are $n_\text{D} \times n_\text{D}$ matrices satisfying the Clifford algebra, $\{ \gamma^{\mu},\gamma^{\nu} \} = 2 \delta^{\mu \nu}$. In these expressions, we have set the speed of light to unity. In $D=3$, a Dirac spinor has a minimum of $n_\text{D}=2$ components; however, in applications to condensed matter systems, the number of two-component Dirac fermions in a bulk lattice system is always doubled due to fermion doubling arguments~\cite{Vafek2013,Nielsen1981}, so the total number of fermionic degrees of freedom $N$ is always a multiple of four.

In $D=3$, there is a critical value of the tuning parameter, $s = s_c$, such that for ${s < s_c}$ the scalar order parameter acquires a finite expectation value, ${\langle \phi \rangle \neq 0}$. Such a finite expectation value spontaneously breaks the ($\mathbb{Z}_2$) parity symmetry of the theory, which is given by taking $(x^0,x^1,x^2) \rightarrow (x^0,-x^1,x^2)$ together with
\beq
	\Psi \rightarrow \gamma^1 \Psi\,, \quad \overline{\Psi} \rightarrow -\overline{\Psi} \gamma^1\,, \quad \phi \rightarrow -\phi\,.
	\label{eq:GNYparity}
\eeq
The finite expectation value of $\phi$ acts as a Dirac mass in \eqref{eq:LGNY}, resulting in a massive spectrum of fermions above a two-fold degenerate ground state. In contrast, for ${s > s_c}$, the parameter $s$ flows to positive infinity while $g_{\mathrm{Y}}$ and $\lambda$ flow to zero. In this limit, we may ignore the gapped bosonic fields, and at long distances the theory describes a SM of non-interacting, massless Dirac fermions with the Euclidean Lagrangian
\beq
	\mathcal{L}_\text{D} = - \overline{\Psi}^{j} \cancel\partial \Psi^{j} .
	\label{eq:Ldirac}
\eeq
We call this fixed point the Dirac CFT, and its properties are easily obtained since $\mathcal{L}_\text{D}$ is exactly solvable.

Directly at the QCP, $s=s_c$, the interaction couplings $g_{\rm Y}$ and $\lambda$ flow to non-zero values of an  interacting fixed point, determined by the chiral Ising universality class. We hence  denote the critical theory of this emerging  interacting fixed point  the chiral Ising CFT. This QCP is non-perturbative directly in $D=3$, but there exists a perturbative expansion in $\epsilon = 4 - D$, where $\lambda \sim g_{\rm Y}^2 \sim O(\epsilon)$, and the universal properties of the QCP may be obtained after extrapolating to $\epsilon = 1$. This will be our primary analytic tool for studying the finite-size torus spectrum as detailed in Sec.~\ref{sec:eps_expansion}.

One may alternatively describe the above QCP using a purely fermionic field theory, the Gross-Neveu (GN) model~\cite{Gross1974}, whose imaginary-time Lagrangian is 
\begin{equation}
    \mathcal{L}_\text{GN} = - \overline{\Psi}^j \cancel\partial \Psi^j - \frac{g}{2} \left( \overline{\Psi}^j \Psi^j \right)^2,
    \label{eq:LGN}
\end{equation}
with a self-interaction of strength $g>0$. For $D=3$, the coupling $g$ is renormalization group (RG) irrelevant in perturbation theory, so a weak-coupling analysis always results in a stable massless Dirac SM phase with $\langle \overline{\Psi} \Psi \rangle = 0$. However, there is ample evidence for a non-perturbative UV fixed point at some value $g = g_c$, where for $g>g_c$ the system flows to strong coupling. At strong coupling, the system dynamically generates a mass by acquiring an expectation value $\langle \overline{\Psi} \Psi \rangle \neq 0$, spontaneously breaking the ($\mathbb{Z}_2$) parity transformation of \eqref{eq:GNYparity}, which will be examined in more detail in Sec.~\ref{sec:symmetry}. By the principle of universality, this fixed point should also be in the chiral Ising universality class, although it is only analytically accessible in an expansion in $\epsilon' =D-2$~\cite{zinn1991}.

We note that both the GNY and GN models may be studied directly in $D=3$ (and even in fractional dimensions $2<D<4$) by a perturbative expansion in $1/N$, where it may be shown that the fixed points of the two models are exactly equivalent in the scaling limit within this expansion \cite{zinn1991}. We discuss the $1/N$ and $\epsilon'$ expansions of the torus spectrum in Appendix \ref{app:largen}, where we additionally give checks that the three expansions all give consistent torus spectra to leading order.

In both the GNY and GN field theories, there is a global U($N_f$) symmetry obtained by taking $\Psi^j \rightarrow U^{j j'}\Psi^{j'}$, with $U \in \mathrm{U}(N_f)$. When $n_\text{D} = 2$, it is sometimes conventional to decompose each Dirac fermion into two Majorana fields, after which the theory is invariant under the larger group O($2N_f$). This Majorana formulation is conventionally used to define the O($N$)-invariant chiral Ising CFTs \cite{Iliesiu2015,Iliesiu2018}. For $n_\text{D} > 2$, these field theories no longer have an explicit O($N$) symmetry; however, due to the structure of the perturbative expansion for the beta functions, the scaling dimensions of all operators turn out to only depend on $N$ rather than $n_\text{D}$ or $N_f$ separately, which leads to identical critical properties for all theories with the same total number of components $N = n_\text{D} N_f$. We will show that the torus spectrum also only depends on $N$. Therefore, we conjecture that all of the chiral Ising CFTs with the same number of total degrees of freedom $N$ flow to the same O($N$)-invariant CFTs irrespective of the smaller global symmetries present in the Lagrangians of Eqns.~(\ref{eq:LGNY}) and (\ref{eq:LGN}), and will also obey an identical critical torus energy spectrum.

This emergent O($N$) symmetry also results in a particular degeneracy structure of the low-energy eigenstates of the critical torus energy spectrum, as shown in Secs.~\ref{sec:gny_cft} and~\ref{sec:eps_expansion_structure}.

\subsection{Model Hamiltonians}
\label{sec:model_ham}

Our perturbative analysis of the GNY torus spectrum will be complemented by the non-perturbative analysis of microscopic fermionic quantum lattice models that exhibit a QCP which is widely believed to belong to the chiral Ising universality class as described by the GNY field theory~\eqref{eq:LGNY}.  In particular, we consider two  models of spinless fermions with Dirac cones in the non-interacting limit at half filling. 

\begin{figure}[t]
	\centering
	\includegraphics[width=\columnwidth]{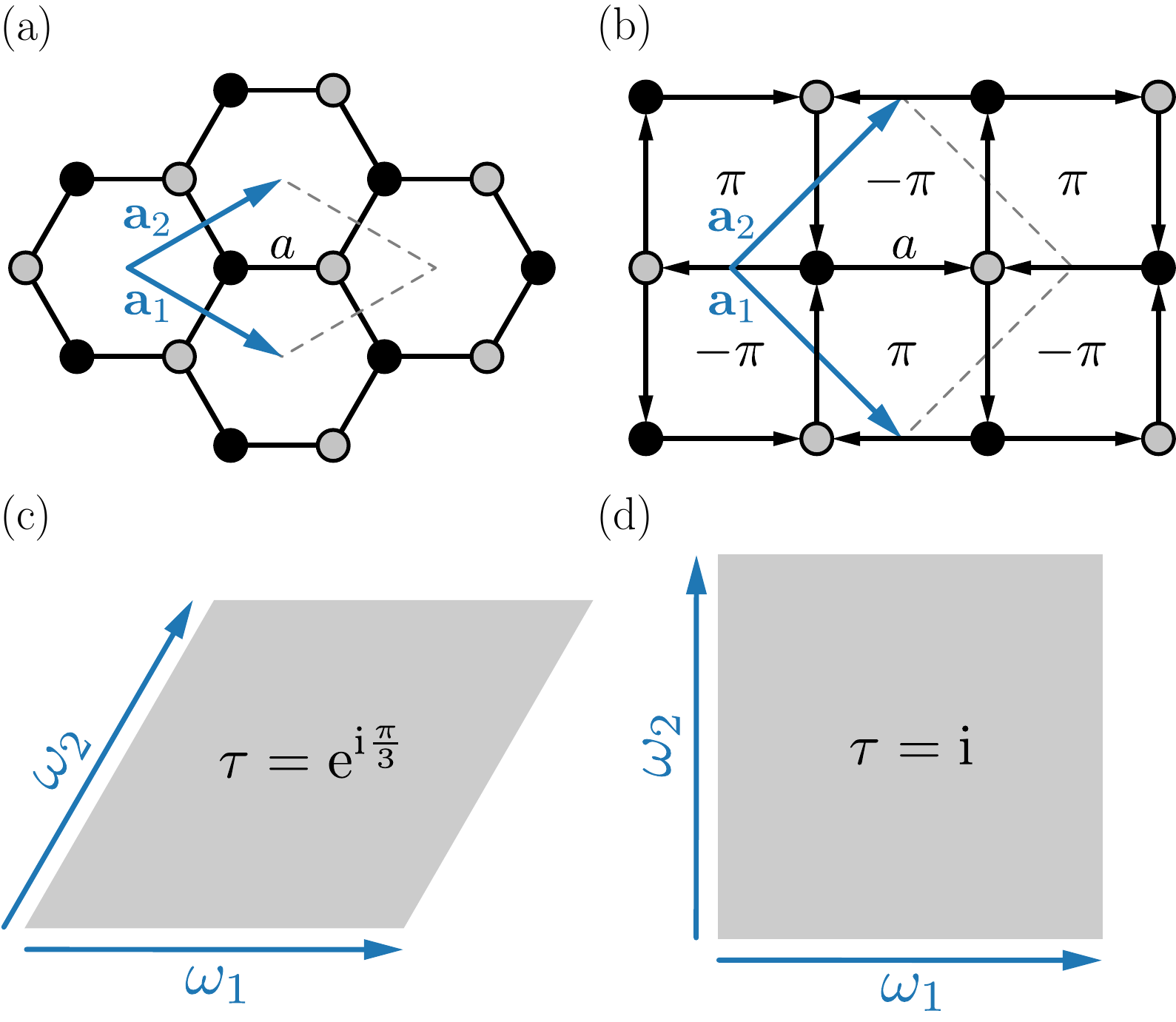}
	\caption{Illustration of (a) the honeycomb lattice, (b) the $\pi$-flux square lattice, and the two corresponding torus geometries with (c) sixfold, and (d) fourfold rotational symmetry. The two sublattices $A$ and $B$ are indicated by black and gray points respectively. The lattice constant $a$ is given by the distance between nearest neighbors, the lattice vectors $\mathbf{a}_1$ and $\mathbf{a}_2$ are indicated by the blue arrows, and the unit cell is traced by the grey dotted lines. Finite clusters with $N_s = 2 L^2$ sites span $L$ unit cells in the direction of $\mathbf{a}_1$ and $\mathbf{a}_2$ respectively, i.e., $|\omega_1| = |\mathbf{a}_1| L$, $|\omega_2| = |\mathbf{a}_2|L$. In (b) we have chosen the specific gauge $\theta_{ij}=\pi/4$, $\theta_{ji} = -\theta_{ij}$, where the tail (head) of the arrows indicates site $i$ ($j$).}
	\label{fig:lattice}
\end{figure}

\begin{figure}[t]
	\centering
	\includegraphics[width=\columnwidth]{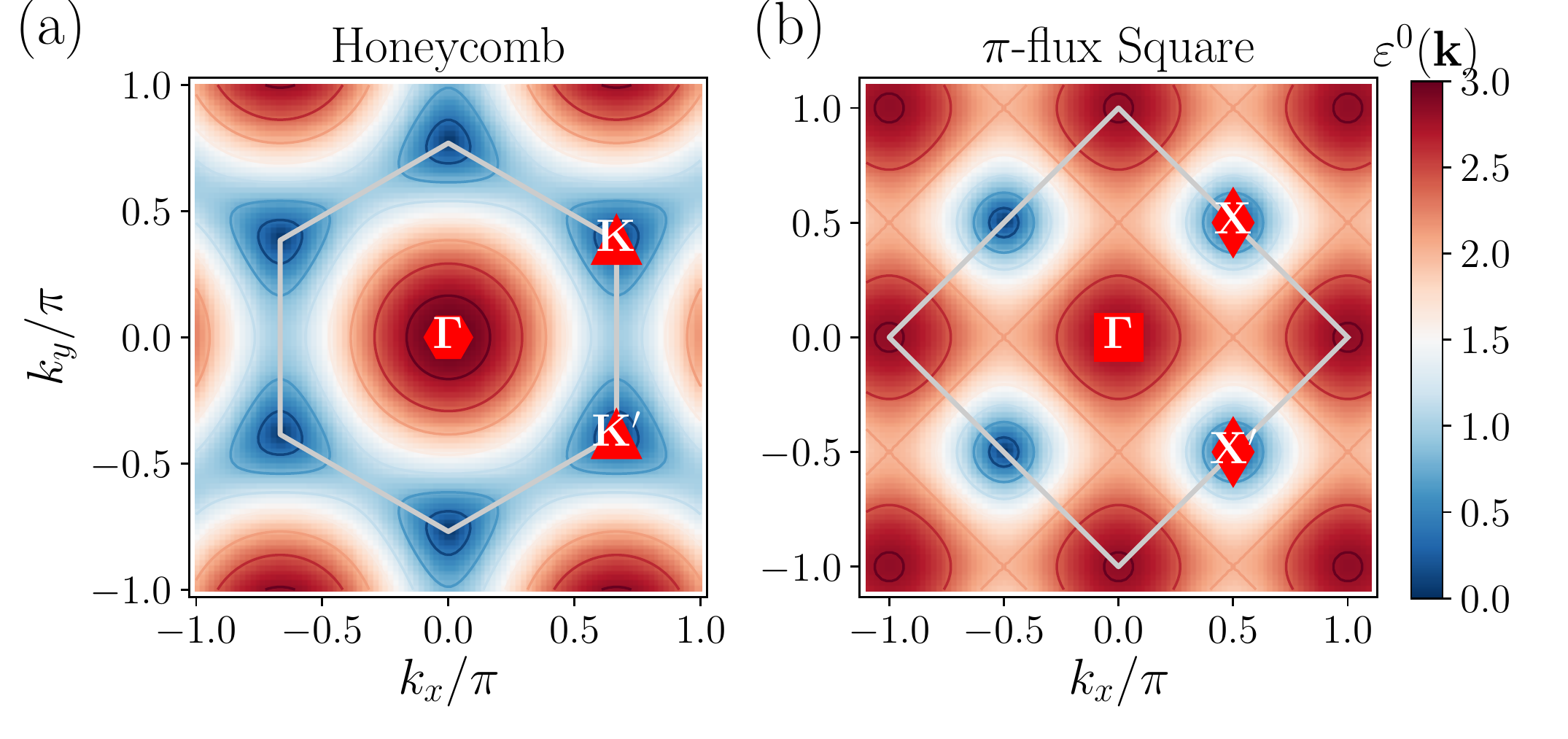}
	\caption{Tight-binding dispersion relation (upper band $\varepsilon^0(\mathbf{k})\geq0$) on the (a) honeycomb lattice, (b) $\pi$-flux square lattice. Dirac points are found at two distinct points in the BZ labelled (a) $\mathbf{K}$ and $\mathbf{K}^\prime$ and (b) $\mathbf{X}$ and $\mathbf{X}^\prime$, respectively. The grey line traces the BZ boundary.}
	\label{fig:gnymodels}
\end{figure}

The first model is defined on a honeycomb lattice [see \figref{fig:lattice}(a)] with the Hamiltonian 
\begin{equation}
    H_{\rm h} = -t \sum_{\nnb} \left(c_i^\dagger c^{\phantom{\dagger}}_j + \text{h.c.} \right) + V \sum_{\nnb} \left(n_i - \frac{1}{2} \right)
    \left(n_j - \frac{1}{2} \right).
    \label{eq:hcmodel}
\end{equation}
Here, $\nnb$ stands for nearest neighbour bonds on the honeycomb lattice, $c_i^{(\dagger)}$ denotes fermionic annihilation (creation) operators at site $i$, and $n_i = c_i^\dagger c^{\phantom{\dagger}}_i$ is the fermionic number operator. The first term in \eqref{eq:hcmodel} is the tight-binding model description of  the fermionic hopping between nearest neighbour sites. The free dispersion relation for positive frequencies is shown in \figref{fig:gnymodels}(a) and has Dirac cones at the two non-equivalent Dirac points in the Brillouin zone (BZ), denoted $\mathbf{K}$ and $\mathbf{K}'$, characteristic for a SM state. At half-filling the spectrum is particle-hole symmetric. The second term describes a density-density interaction between fermions on neighboring sites, driving the system, for $V\gg t$, into a charge density wave (CDW) state, in which the particle-hole symmetry, together with a sublattice exchange parity symmetry, (a $\mathbb{Z}_2$ symmetry group) is spontaneously broken and the local densities within the two sublattices differ in the thermodynamic limit. The position of the QCP was determined previously to be at $V_c/t \approx 1.355$~\cite{Wang2014,Li2015,Hesselmann2016,Wang2016}.

The second model that  we  consider describes interacting, spinless fermions on the $\pi$-flux square lattice at half filling [see \figref{fig:lattice}(b)], with the Hamiltonian
\begin{equation}
    H_{\rm s} = -t \sum_{\nnb} \left( \E^{\I \theta_{ij}} c_i^\dagger c^{\phantom{\dagger}}_j + \text{h.c.} \right) + V \sum_{\nnb} \left(n_i -
    \frac{1}{2} \right) \!\!\left(n_j - \frac{1}{2} \right).
    \label{eq:sqmodel}
\end{equation}
While the phases $\theta_{ij}$ depend on the choice of gauge, the total flux per plaquette, $\phi = \sum_{\nnb \in \Box} \theta_{ij}$ is fixed to be $\phi= \pi$. In the non-interacting case, $V=0$, \eqref{eq:sqmodel} is a tight-binding model with, in general, complex hopping amplitudes. Its dispersion relation (for the specific choice of gauge $\theta_{ij} = \pi/4$) is plotted in \figref{fig:gnymodels}(b) and again shows two distinct Dirac cones (now at the $\mathbf{X}$ and $\mathbf{X}'$ points) in the BZ. For large values of $V\gg t$,  this model also exhibits a CDW phase with different densities on the two sublattices that are coupled by the repulsion term. The position of the QCP was estimated to be at $V_c/t \approx 1.279$ \cite{Wang2014, Li2015, Huffman2017}. In the remainder of this paper, we will set the energy scale of the lattice models by fixing the hopping amplitude $t=1$.

The QPTs in both lattice models introduced above may be described by either the GNY or GN quantum field theories with a single ($N_f=1$) flavor of $n_\text{D}=4$-component spinors~\cite{Herbut2006, Herbut2009}, corresponding to a value of $N=4$, as reviewed in Appendix~\ref{app:micromap}. If not  specified otherwise, we mean for $N$ to take on this value in the following.  An important aspect of the correspondence between the microscopic lattice models and the effective QFT description is the manifestation of several global symmetries, which we will examine in the following section. 

\subsection{Symmetries}
\label{sec:symmetry}

The model Hamiltonians $H_{\rm h}$ and $H_{\rm s}$ possess several global symmetries, some of which are spontaneously broken in the ordered phase. Here we review these relevant symmetries and discuss how they manifest in the GNY and GN field theories.

Since both model Hamiltonians are  bipartite, we may label the fermion annihilation (creation) operators as ${c^{(\dagger)}_{\alpha}(x_i,y_i)}$, where ${\mathbf{r}_i = (x_i,y_i)}$ is the coordinate of the Bravais lattice, and $\alpha = A,B$ is the sublattice index. In this section, we take the convention from \figref{fig:lattice} that the $A$ and $B$ sites are connected along the $x$-axis, and that the unit cell is centered on the point directly equidistant between these two sites. Furthermore, with regards to $H_{\rm s}$, we consider here, for simplicity, a gauge, in which the phases $\E^{\I \theta_{ij}}$ in $H_{\rm s}$ are all real (for example, by choosing $\theta_{ij} = \pi$ on one link of each square plaquette and zero on the others); then these symmetries take on an identical form in both models. With these conventions, we can now describe the symmetries of both models. 

First, we have the usual global U(1) symmetry from particle-number conservation, $c_{\alpha}(x_i,y_i) \rightarrow \E^{\I \varphi} c_{\alpha}(x_i,y_i)$. In addition, both models have an anti-unitary time-reversal symmetry that is given by complex conjugation in real space, leaving the fermionic operators unchanged,
\beq
	T: \qquad c_{\alpha}(x_i,y_i) \rightarrow c_{\alpha}(x_i,y_i).
\eeq
We may also define a parity flip across either the vertical or horizontal axes, $I_x: (x_i,y_i) \rightarrow (-x_i,y_i)$ and  $I_y: (x_i,y_i) \rightarrow (x_i,-y_i)$. These parity symmetries are actually part of the larger point group symmetries of these models, e.g., the dihedral symmetry group $D_6$ on the honeycomb lattice. Besides a change in coordinates, our convention of parity implies that $I_x$ also exchanges the two sublattices, so we have
\beq
	I_x: \qquad \begin{pmatrix}
		c_{A}(x_i,y_i) \\
		c_{B}(x_i,y_i)
	\end{pmatrix} \rightarrow 
	\begin{pmatrix}
		c_{B}(-x_i,y_i) \\
		c_{A}(-x_i,y_i)
	\end{pmatrix},
\eeq
\beq
	I_y: \qquad \begin{pmatrix}
		c_{A}(x_i,y_i) \\
		c_{B}(x_i,y_i)
	\end{pmatrix} \rightarrow 
	\begin{pmatrix}
		c_{A}(x_i,-y_i) \\
		c_{B}(x_i,-y_i)
	\end{pmatrix}.
\eeq
Finally, we also define a particle-hole transformation,
\beq
	\mathcal{C}: \qquad \begin{pmatrix}
		c_{A}(x_i,y_i) \\
		c_{B}(x_i,y_i)
	\end{pmatrix} \rightarrow 
	\begin{pmatrix}
		c^{\dagger}_{A}(x_i,y_i) \\
		-c^{\dagger}_{B}(x_i,y_i)
	\end{pmatrix},
\eeq 
under which the  density on each site transforms as ${n_i \rightarrow (1 - n_i)}$. In the CDW phase, the densities of fermions on the sublattices $A$ and $B$  differ, such that both $I_x$ and $\mathcal{C}$ are spontaneously broken.
 
We now characterize the symmetries of the field theory. As shown in Appendix \ref{app:micromap}, the GNY and GN field theories may be derived from our model Hamiltonians in the continuum and scaling limits, and we find that the models possess a single flavor of four-component Dirac fermions, i.e., $n_\text{D}=4$, $N_f = 1$. In discussing this realization of the field theory, it is useful to introduce the following explicit representation of the gamma matrices, which arises naturally from the derivation:
\beq
	\gamma^0 = \begin{pmatrix}
		\sigma^z & 0 \\
		0 & \sigma^z
	\end{pmatrix}, \quad
	\gamma^1 = \begin{pmatrix}
		\sigma^x & 0 \\
		0 & \sigma^x
	\end{pmatrix}, \quad
	\gamma^2 = \begin{pmatrix}
		\sigma^y & 0 \\
		0 & -\sigma^y
	\end{pmatrix}.
	\label{eq:gamma_rep}
\eeq
In this representation, the first two indices of the four-spinors represent fermions on the honeycomb (square) lattice with momentum near $\mathbf{K}$ ($\mathbf{X}$) at sublattice A and B respectively, while the bottom two components represent fermions near momentum $\mathbf{K}'$ ($\mathbf{X}'$) at sublattice A and B. It is useful to define two additional gamma matrices, which anticommute with the above,
\beq
	\gamma^3 = \begin{pmatrix}
		0 & \sigma^y \\
		\sigma^y & 0
	\end{pmatrix}, \quad
	\gamma^4 = \begin{pmatrix}
		0 & -\I \sigma^y \\
		\I \sigma^y & 0
	\end{pmatrix}.
	\label{eq:gamma_extra}
\eeq
We also define ten Hermitian matrices $\gamma^{a b} \equiv \I \gamma^{a} \gamma^{b}$, where $0 \leq a < b \leq 4$. This parametrization is useful because an arbitrary $4 \times 4$ Hermitian matrix may be written as a linear combination of the sixteen matrices $\{ \mathbb{I}, \gamma^{a}, \gamma^{a b} \}$ with real coefficients.

Given this explicit representation, we may follow the derivation of Appendix \ref{app:micromap} to obtain how the microscopic symmetry transformations act on the fields of the QFT. In particular, we use the fact that the transformations $T$, $I_y$, and $\mathcal{C}$ exchange the Dirac points, while $I_x$ does not (see also Fig.~\ref{fig:gnymodels}). The U(1) symmetry simply takes the form $\Psi \rightarrow \E^{\I \varphi} \Psi$, while the discrete symmetries are given by
\bea
	T:& \quad \ \ \ \Psi \rightarrow \gamma^{24}  \Psi, \qquad &\phi \rightarrow \phi, \nn
	I_x:& \quad \Psi(x,y) \rightarrow \gamma^1 \Psi(-x,y), \quad &\phi(x,y) \rightarrow - \phi(-x,y), \nn
	I_y:& \quad \ \Psi(x,y) \rightarrow  \gamma^{24} \Psi(x,-y), \quad &\phi(x,y) \rightarrow \phi(x,-y), \nn
	\mathcal{C}:& \quad \quad \ \ \Psi \rightarrow \left( \Psi^{\dagger} \gamma^{13} \right)^{\mathrm{T}}, \quad &\phi \rightarrow -\phi,
\eea
where $T$ is anti-unitary. It is straightforward to show that Eqns.~(\ref{eq:LGNY}) and (\ref{eq:LGN}) are invariant under these transformations (where we simply ignore the transformation rules on $\phi$ in the GN case). We furthermore see that the order parameters for the $\mathbb{Z}_2$ symmetry breaking may be given by either $\langle \phi \rangle$ or $\langle \overline{\Psi} \Psi \rangle$, which are both odd under $I_x$ and $\mathcal{C}$ but even under the rest of the above symmetry transformations.

In addition to the symmetries inherited from the microscopic model, the field theory has additional symmetries which are not present in the lattice model. Importantly, it is invariant under the three-dimensional Lorentz group, which is generated by $\{\gamma^{01},\gamma^{02},\gamma^{12}\}$~\footnote{In application to the imaginary-time Euclidean theories, these are SO(3) rotations, but they will correspond to SO(2,1) Lorentz transformations after a Wick rotation to real-time.}, and includes the discrete rotational symmetries of the lattice as a subgroup. Finally, we have an U(1) (``chiral'') symmetry given by $\Psi \rightarrow \exp(\I \gamma^{34}) \Psi$, which corresponds to performing independent U(1) rotations at the two inequivalent Dirac points. We stress that these emergent symmetries of the field theory only apply to the lattice models in the strict scaling limit, and any degeneracies found in the torus spectrum of the field theory due to these particular symmetries are expected to be approximate in the lattice models due to the presence of additional irrelevant operators.

\subsection{Torus Compactifications}
\label{sec:torus}

In  numerical studies of two-dimensional quantum lattice models, one typically considers finite-size clusters constructed from the underlying lattice,  with periodic boundary conditions taken in both lattice directions. This way, one effectively studies a torus compactification of the original infinite lattice  model. Here, we choose finite-size clusters which preserve the maximal six- (four-)fold rotational symmetry $C_6$ ($C_4$) of the honeycomb (square) lattice model. As  mentioned in the introduction, our analysis serves the dual purpose of (i) examining the  energy level structure of  fermionic model systems on such torus geometries, which serves as a universal fingerprint for the corresponding QFT, as well as (ii) deriving appropriate estimators for other physical quantities, such as the effective Fermi velocity (the effective speed of light) of the interacting fermion models, which depend on  such spectroscopic data. 

\begin{figure}[t]
	\centering
	\includegraphics[width=\linewidth]{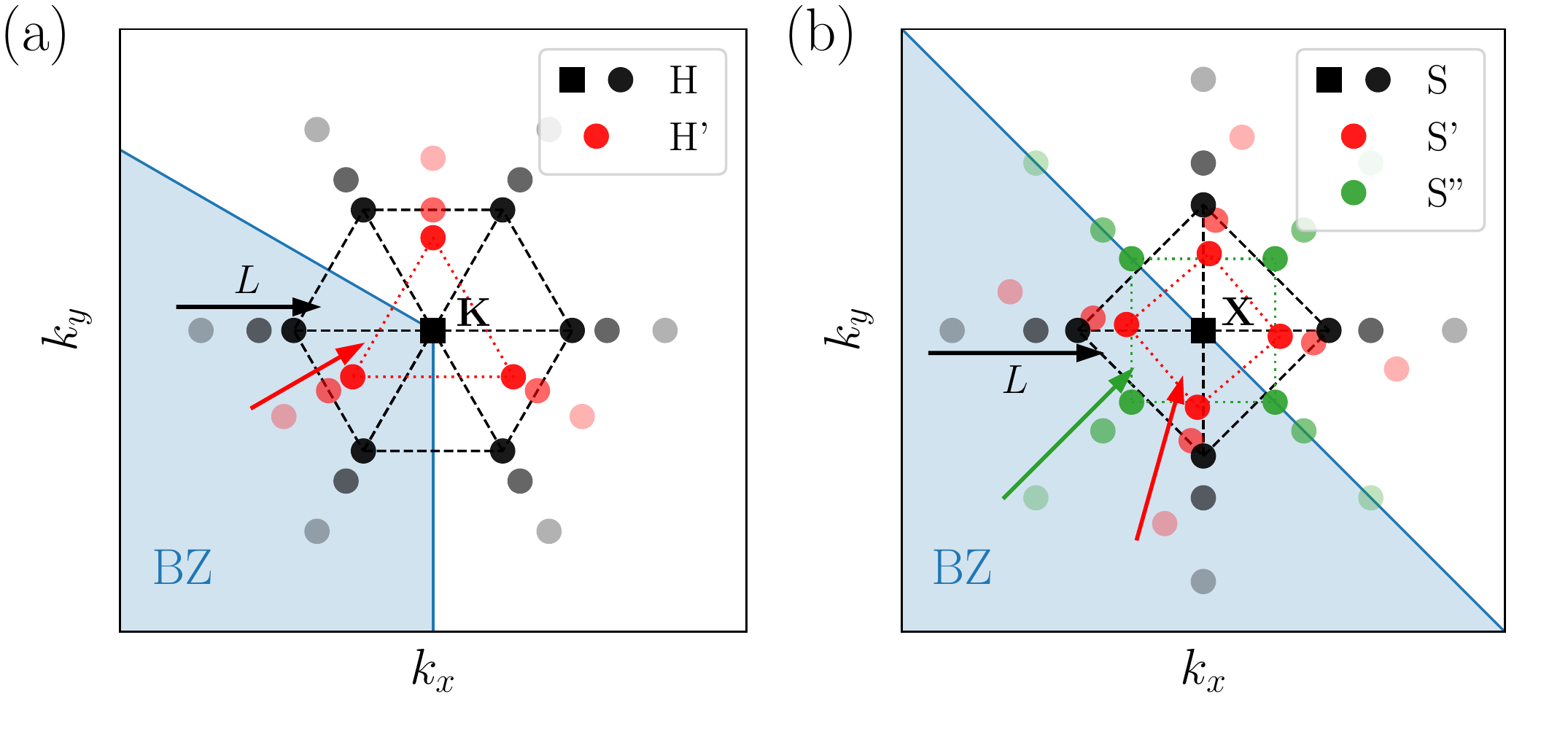}
	\caption{Illustration of the different finite-size cluster families for the (a) honeycomb, (b) square lattice. Shown is a zoom into the momentum space around one of the Dirac points $\mathbf{K} / \mathbf{X}$. The closest momentum points are plotted for clusters of different linear size $L$, and, as a guide to the eye, we have connected them by lines for the largest $L$ shown. The Dirac points are (not) part of the momentum space for cluster families H/S (H'/S'/S''). See text for details.}
	\label{fig:lattice_families}
\end{figure}
 
The torus clusters for the  two microscopic models that we consider here exhibit different overall rhombic shapes [see~\figref{fig:lattice}(c), (d)], which act as an infrared (IR) cutoff (irrespectively of the lattice discretization, i.e., the ultraviolet cutoff). The IR characteristics remain influential in the thermodynamic limit~\cite{Schuler2016}, and we account for them also in the analysis of the GNY field theory on finite tori. This is done as follows: In the continuum limit, one can use complex coordinates on the two-dimensional torus, $x = x_1 + \I \, x_2$. The torus is then defined by two complex periods, $\omega_1$ and $\omega_2$, such that the points $x + n \omega_1 + m \omega_2$ are equivalent for all $n,m \in \mathbb{Z}$. The torus shape is then characterized by its modular parameter, $\tau \equiv \omega_2/\omega_1 = \tau_1 + \I \tau_2$. In particular, the considered triangular-lattice based tori (such as for the honeycomb lattice model) have a value of $\tau = \exp(\I \pi/3)$, while the square-lattice based tori correspond to $\tau = \I$, respectively, \mbox{c.f.~Fig.~\ref{fig:lattice} (c), (d)}. In the framework of the GNY field theory the torus corresponds to periodic boundary conditions for both the $\Psi$ and the $\phi$ fields.

The lattice models we consider all use periodic boundary conditions for their simulation clusters. Since the Dirac points in the considered models are not located at the $\Gamma$ point ($\mathbf{k}=(0, 0)$), several families of clusters arise, which differ in their momentum discretization grid around the Dirac points, as illustrated in Fig.~\ref{fig:lattice_families}. 
For example, when considering finite torus clusters with $N_s=2 L^2$ lattice sites of the honeycomb lattice (c.f. \figref{fig:lattice}), those with linear size $L\!\!\mod3=0$ feature the Dirac points ($\mathbf{K}$ and $\mathbf{K}'$) in their momentum space (family H), in contrast to clusters with $L\!\!\mod3\neq0$, which do not feature the Dirac points (family H'), so that the spectrum is gapped already in the tight-binding limit, $V=0$. Hence, the finite-size torus spectrum is qualitatively different for those two families already in the non-interacting limit, and we observe characteristic differences also for the interacting case. The case \mbox{$L\!\!\mod3=0$} in the lattice models corresponds to the standard periodic boundary condition case in the GNY field theory discussed above. The second case $L\!\!\mod3\neq0$ corresponds to the GNY field theory with twisted boundary conditions, $\Psi(x + n \omega_1 + m \omega_2) = e^{i \theta} \Psi(x)$. We will present numerical results for those spectra later on, although we will only give a few comments about the structure of the $\epsilon$-expansion for twisted boundary conditions in Section \ref{sec:eps_expansion}. Similar considerations apply to the fermionic model $H_{\rm s}$ on the $\pi$-flux square lattice, as detailed in Sec.~\ref{sec:hc_numerics_spec}.  

\begin{figure*}[ht]
	\centering
	\includegraphics[width=\linewidth]{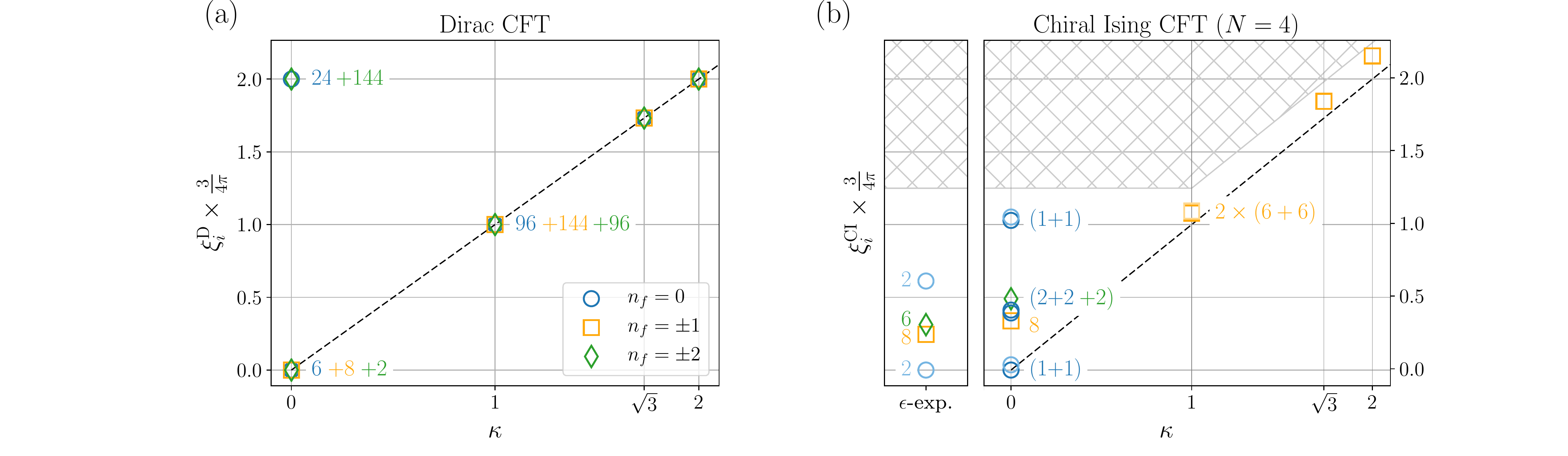}
	\caption{Critical torus low-energy spectrum for (a) the free, massless Dirac CFT, and (b) the $N=4$-component chiral Ising CFT, as a function of the reduced momentum $\kappa$ on honeycomb tori  that contain the Dirac points (family H). Different symbols and colors indicate the fermion number sectors $n_f$ relative to half filling. The numbers in the plots indicate the degeneracy of the levels, while the parenthesis in (b) indicate  nearly degenerate levels as described in the main text. The spectrum in (a) is analytically exact, the spectrum in the right panel of (b) is based on extrapolated numerical data from the microscopic lattice model $H_{\rm h}$. The left panel in (b) shows results from the $\epsilon$-expansion for $\tau=\text{exp}(\I \pi/3)$ and $\kappa=0$ as a comparison. The black, dashed lines show a linear dispersion according to the Fermi velocity $v^0_\text{F}=3/2$ and $v^c_\text{F}$ in the panels (a) and (b), respectively. Here, $v^c_\text{F}$ has been estimated as described in Sec.~\ref{sec:vf_ren}. The hatching in (b) indicates that we have not computed energy levels within this higher energy regime.}
	\label{fig:spec_vs_kappa}
\end{figure*}

\section{Overview of the Central Results}
\label{sec:overview}

This section provides an overview of our analytical and numerical findings. Further details are provided in the subsequent sections of this paper. For all the models and the field theories that we consider, the phase diagram is divided by a strongly coupled QCP (chiral Ising CFT) into an extended SM regime of massless Dirac fermions (Dirac CFT) and a regime with spontaneous symmetry breaking and gapped fermionic excitations. While the spectroscopic properties of the strongly coupled chiral Ising fixed point are of particular interest, the excitation spectrum of the free Dirac CFT characterizes the SM regime, which thus exhibits distinctly different spectral characteristics. Additionally, there are important crossover effects between these two regimes which need to be treated with care. We therefore start by exploring the spectroscopic properties of the Dirac CFT, before discussing our main findings for the chiral Ising CFT. We then discuss the crossover between these two CFTs as well as the subtleties in obtaining the correct Fermi velocity renormalization. In this section, we concentrate on the case of the honeycomb lattice model $H_{\rm h}$ and torus clusters that contain the Dirac points (family H).

\subsection{Dirac CFT Torus Spectrum}
\label{sec:dirac_cft}

In the SM phase with $V < V_c$ the torus spectrum is characterized by the free, massless Dirac CFT, defined by Eq. (\ref{eq:Ldirac}). Its excitation energies can be readily calculated analytically, and they are directly related to the Fermi velocity of the Dirac fermions. For large finite clusters with $N_s=2L^2$ lattice sites the energy levels scale as $1/L$, and the torus spectrum in the SM regime is then given by
\begin{equation}
    \Delta_i = \frac{v_\text{F}(V)}{L} \xi^\text{D}_i,
    \label{eq:gap_dirac}
\end{equation}
where $\Delta_i $, $i=1,2,...$ denotes the set of finite-size energy gaps (relative to the ground state energy) which make up the low-energy torus spectrum. Here, $v_\text{F}(V)$ is the renormalized Fermi velocity at interaction strength $V$, and the $\xi^\text{D}_i$ is a set of universal numbers that characterize the Dirac CFT~\cite{Schuler2016, Whitsitt2016, Whitsitt2017}. The values $\xi_i^\text{D}$ in the low-energy regime are shown in \figref{fig:spec_vs_kappa}(a). Additional quantum numbers are attached to the corresponding eigenstate, in particular  the fermion number $n_f$ relative to half filling, and its momentum $\mathbf{k}$. At half-filling, many-body states of vanishing finite-size gaps reside  both at the Dirac points $\mathbf{K}$ and $\mathbf{K}'$, as well as at total momentum $\mathbf{k}=0$~\footnote{Note that in the field theory the soft energy levels are around $\mathbf{k}=0$, as the information about $\mathbf{K}$ and $\mathbf{K}'$ has been absorbed into the spinor components.}. For example, at $n_f=0$ we have to put two fermions into the upper or lower band of the Dirac cones at $\mathbf{K}$ and $\mathbf{K}'$ to find in total six zero energy states. This can be done by either putting both fermions into the two-bands of a single Dirac cone which results in a total momentum of $\mathbf{k} = 2 \mathbf{K}^{(\prime)} \equiv -\mathbf{K}^{(\prime)}$ of the many-body state, or by putting each one in a different Dirac cone with a total momentum of $\mathbf{k} = \mathbf{K} + \mathbf{K}' \equiv 0$. We thus introduce a reduced momentum variable 
\begin{equation}
    \kappa=\frac{3}{4\pi} L \, |\mathbf{k}\!\!\!\mod \mathbf{K}^{(\prime)}|\,, 
\end{equation}
taken modulo the momentum space lattice vectors spanned by $\mathbf{K}$ and $\mathbf{K}'$. States with small value of $\kappa$ then map to the low-energy sector described by the QFT. The normalization factor $3/(4\pi) L$ is chosen such that momenta closest to $\mathbf{k}=0$ and to the Dirac points correspond to a value of $\kappa=1$ (for lattice constant $a=1$). The levels $\Delta_i$ together with their quantum numbers and multiplicities provide a characteristic fingerprint of the Dirac CFT. We show the low-energy part of the  torus spectrum of the  Dirac CFT as a function of the reduced momentum in \figref{fig:spec_vs_kappa}(a), which also displays the corresponding multiplicities of each level. 

In the following, we will refer to gaps in the torus spectrum as denoting finite differences between the rescaled energy gaps ${L \Delta_i \propto \xi_i}$. For example, in \figref{fig:spec_vs_kappa}(a), the lowest level at $\kappa=1$ has a finite gap with respect to the lowest level at $\kappa=0$. The raw many body spectrum in the thermodynamic limit, however, shows no gaps for ${V \leq V_c}$. This notion of gaps in the torus spectrum will turn out particularly useful to quantify the differences between the torus spectrum of the Dirac CFT and the one at the chiral Ising critical point, which we discuss next.

\subsection{Chiral Ising  CFT Torus Spectrum}
\label{sec:gny_cft}

At the critical point $s=s_c$, or $V=V_c$, our models are described by the strongly interacting chiral Ising CFT, for which no exact analytical solutions are known. An analytical approach to the critical torus spectrum of the chiral Ising fixed point of the $N=4$ component GNY field theory is provided by the $\epsilon$ expansion as detailed in Sec.~\ref{sec:eps_expansion}. From these calculations and numerical simulations of the microscopic models, we find that the critical torus spectrum of the interacting chiral Ising CFT [i.e., at finite $g_{\rm Y}>0$ in \eqref{eq:LGNY}] is characterized by a (different) set of finite-size energy gaps $\Delta_i$ that scale as
\begin{equation}
    \Delta_i = \frac{v^c_\text{F}}{L} \xi^\text{CI}_i\,,
    \label{eq:gap_gny}
\end{equation}
with $v^c_\text{F}=v_\text{F}(V_c)$ the renormalized Fermi velocity at the critical interaction strength. Such a scaling form of the critical torus spectrum of an interacting fixed point has been obtained also in studies of purely bosonic quantum critical points~\cite{Schuler2016, Whitsitt2016, Whitsitt2017}. It can be considered  a mass spectrum of the quantum critical theory with a mass scale set by the IR cutoff, which is proportional to $v^c_\text{F}/L$. Here, the $\xi^\text{CI}_i$ are again a set of universal numbers which are, however, distinct from those of the  Dirac CFT, and, together with the levels multiplicities and quantum numbers, identify the chiral Ising CFT. The $\epsilon$-expansion predicts a rich level multiplicity structure for the low-energy levels because of the before mentioned emergent O($N$) symmetry of the chiral Ising field theory.

\begin{figure*}[t]
	\centering
	\includegraphics[width=\linewidth]{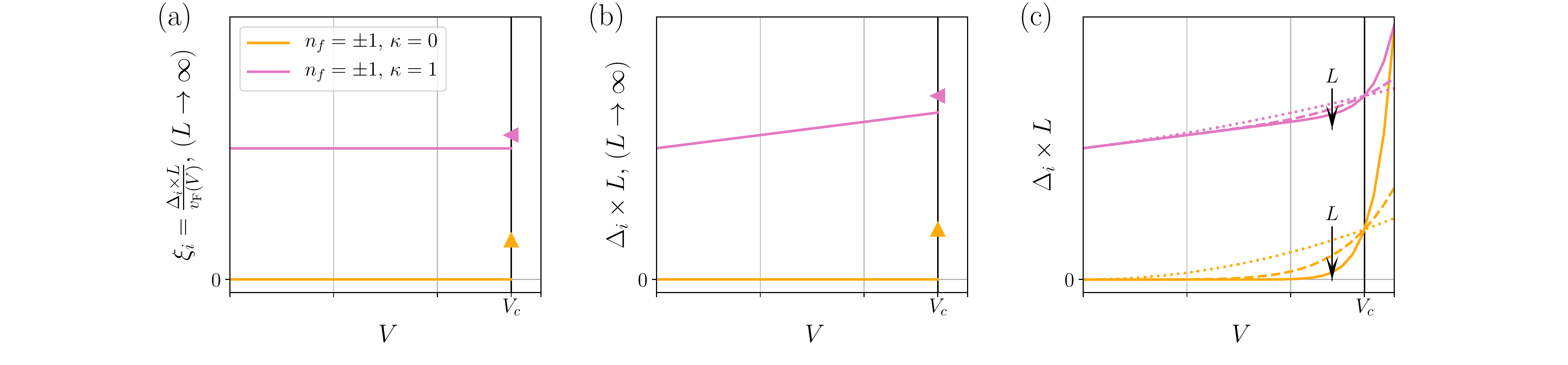}
	\caption{Sketch of crossover effects in the torus spectrum near the chiral Ising CFT. In (a) we show the universal numbers $\xi_i$ for two different levels within the SM phase $V<V_c$ and at the QCP, $V=V_c$. (b) shows the scaled energy gaps $\Delta_i \times L \propto v_\text{F}(V) \xi_i$ in the thermodynamic limit $L\rightarrow\infty$, assuming a linear Fermi velocity renormalization $v_\text{F}(V) \propto V$, while (c) illustrates the scaled energy gaps measured from finite size systems with linear size $L$.} 
	\label{fig:crossover} 
\end{figure*}

In \figref{fig:spec_vs_kappa}(b), we show the low-energy part of the chiral Ising torus spectrum for the $N=4$-component GNY theory, normalized by the Fermi velocity $v_\text{F}^c$ as a function of the reduced momentum $\kappa$. In this figure, we compare our estimates for  the $\xi^\text{CI}_i$ as obtained from the $\epsilon$ expansion at $\kappa=0$ (left panel) with finite-size extrapolated gap data from the microscopic lattice model $H_{\rm h}$ (right panel). The structure of level multiplicities and the quantum numbers of the corresponding eigenstates as obtained from the perturbative $\epsilon$-expansion compare remarkably well to the numerical analysis of the microscopic lattice models (when comparing actual values, one needs to keep in mind, that in the $\epsilon$ expansion they result from a simple extrapolation of only the leading term). It is important to note, that the degenerate levels obtained from the perturbative analysis appear as quasi-degenerate levels in the numerical data, because of corrections to scaling present in any particular lattice implementation of the GNY field theory. A very prominent feature of the chiral Ising torus spectrum is the opening of large gaps in the torus spectrum between the two (quasi-) degenerate ground state levels with $n_f=0$ and the other 14 levels at $\kappa=0$ (with $n_f=0,\pm1,\pm2)$, which all contribute to the ground state manifold in the Dirac CFT. For the chiral Ising CFT, the $n_f=\pm1$ states form an 8-fold degenerate level,
while we expect the 2-fold degenerate $n_f=\pm2$ and the two 2-fold degenerate $n_f=0$ levels to form a 6-fold (quasi-) degenerate level.

The other large degeneracies of the  higher levels in the Dirac CFT similarly split up in the chiral Ising CFT. Furthermore, very characteristically, we observe a very low-lying two-fold, nearly degenerate set of levels with the same quantum numbers as the ground state levels, similar to what was also observed in Wilson-Fisher CFTs~\cite{Schuler2016, Whitsitt2017}. The quasi-degenerate nature of the ground state levels also transfers to the lowest $n_f=\pm1$ levels at $\kappa=1$.  Their energies are pushed to values slightly above the linear dispersion relation with the velocity $v_\text{F}^c$, where our best estimate for $v_\text{F}^c$ was obtained as described in Sec.~\ref{sec:vf_ren}. We expect that such a two-fold quasi-degeneracy appears for all the lowest $n_f=\pm1$ levels at all $\kappa>0$, however, these were inaccessible because of  finite-size restrictions in the numerical calculations. A comparison of the critical torus spectrum for the chiral Ising universality class in \figref{fig:spec_vs_kappa}(b) with the critical torus spectrum for the (Wilson-Fisher) Ising universality class~\cite{Schuler2016, Whitsitt2017} where, in particular, only non-degenerate levels appear, shows the strong influence of the fermionic degrees of freedom on the critical spectrum. This, once more, demonstrates the potential of the critical torus energy spectrum as a useful identifier for universality classes.

\subsection{Crossover Effects near the QCP}
\label{sec:cross}

For all values $V<V_c$ the microscopic models, in the thermodynamic limit, flow towards the Dirac CFT fixed point with massless fermionic excitations featuring a linear light-cone, and the spectrum is defined by the universal numbers $\xi_i^\text{D}$. The energy spectrum on finite clusters, however, is affected by a pronounced crossover effect: This derives from the fact that in the vicinity of the QCP at $V_c$, the RG flow is first attracted towards the chiral Ising CFT fixed point on intermediate length scales, and later crosses over to the asymptotic Dirac CFT fixed point only beyond an increasingly larger length scale $L_c$, which diverges upon approaching the QCP. For $V$ near $V_c$, sufficiently large system sizes $L \gg L_c$ are thus required in order to probe the asymptotic Dirac CFT fixed point. As a result, the values of the scaled excitation gaps $\Delta_i\times L$ exhibit a continuous crossover between the asymptotic values and those at the QCP, in particular for those levels, for which $\xi^\text{D}_i$ and $\xi^\text{CI}_i$ differ notably (in particular levels at $\kappa=0$). This crossover behavior is illustrated in \figref{fig:crossover}.

This may be made precise by using the theory of finite-size scaling. Assume we have a $D$-dimensional CFT perturbed by a single relevant operator, with an associated correlation length exponent $\nu >0$. We also perturb our CFT with any number of irrelevant operators, and call the usual critical exponent governing the leading corrections to scaling $\omega > 0$. The above conditions should describe a typical critical point with a single relevant direction being probed by an experiment or numerical simulation. With these definitions, the finite-size spectrum of the perturbed CFT on the torus with ``speed of light'' $v$ and linear extent $L$ is given by \cite{Campostrini2014}
\bea
	\Delta_i &=& \frac{v}{L} \Big\{ \xi_i\big[ L (V - V_c)^{\nu} \big] + (L/\delta)^{- \omega} \zeta_i\big[ L (V - V_c)^{\nu} \big] \nn
	&& + \ O(1/L) \Big\}.
	\label{eq:spectrum_scaling}
\eea
In this expression, the dimensionless scaling functions $\xi_i[x]$ and $\zeta_i[x]$ are universal up to overall multiplicative factors and a normalization of their arguments. We have included the non-universal length scale $\delta$ associated with the addition of the leading irrelevant operator to the CFT, and we only show the leading non-analytic dependence on $L$.

In any lattice model, the spectrum calculated numerically will contain all of these terms, but in this paper we focus on extracting the constants $\xi_i = \xi_i[0]$, which completely characterize the torus spectrum of the \emph{unperturbed} CFT. Therefore, we are really interested in taking the limit
\bea
	\delta \ll L \ll |V - V_c |^{-\nu}.
	\label{eq:FSSlimit}
\eea
By comparing this limit to \eqref{eq:spectrum_scaling}, we see that the first inequality ensures that only $\xi_i[x]$ contributes, while the second ensures that the limit $\xi_i[x \rightarrow 0]$ is taken. In analyzing numerical or experimental results, where the non-universal scale $\delta$ may be anomalously large or $V - V_c$ cannot be tuned with arbitrary precision, one should always check that these conditions hold \footnote{We additionally note that if $\omega$ happens to be very small numerically, then corrections to scaling will remain important for large system sizes. This does not seem to be a problem for our cases of interest; for the Dirac CFT, $\omega_\text{D} = 1$, while for the chiral Ising CFT, $\omega_\text{CI} \sim 0.8$ at four-loops in the $\epsilon$-expansion \cite{Zerf2017}.}.

Applying this reasoning to the chiral Ising CFT, we find that, even if we tune into the semimetal phase, $V < V_c$, our finite-size spectrum will continue to be that of the chiral Ising CFT provided the linear extent of the torus satisfies $L \ll L_c = |V - V_c |^{-\nu_\text{CI}}$. Alternatively, we may apply this analysis to the Dirac CFT, which does not have any relevant operators. Instead, the coupling $V$ is actually an \emph{irrelevant} perturbation to the Dirac CFT, so the length scale $L_c$ should actually be associated with $\delta$ in \eqref{eq:FSSlimit}, and the Dirac CFT spectrum is obtained when $L \gg L_c$. At intermediate length scales, $L \sim L_c$, the torus spectrum is described by the full function $\xi_i[x]$.

\subsection{Quantifying the Fermi Velocity Renormalization} 
\label{sec:vf_ren}

In the SM phase, the considered model systems feature massless fermionic excitations in the thermodynamic limit, with a linear single particle dispersion and a renormalized Fermi velocity $v_\text{F}(V)$ which depends on the interaction strength in a non-universal manner.  This is due to a RG flow towards the Dirac CFT fixed point for all values $V < V_c$. Only exactly at the QCP will the system finally flow to the chiral Ising CFT fixed point. While the shape of the fermion dispersion relation does not change within the SM phase, the Fermi velocity $v_\text{F}(V)$ can differ greatly from the non-interacting value $v_\text{F}^0$ and is model dependent [see \figref{fig:crossover}(b)].

Since the torus energy spectrum is a universal property for the underlying Dirac CFT, the energy levels are given by $\Delta_i  L = v_\text{F}(V) \xi^\text{D}_i$ within the SM phase, where the $\xi_i^\text{D}$ denote universal numbers describing the SM phase and do not depend on the interaction strength $V<V_c$. Hence, it is  possible to determine the renormalization of the Fermi velocity from single energy levels (with $\xi_i^\text{D}>0$) of the spectrum,
\begin{align}
	v_\text{F}(V) =   \lim_{L \rightarrow \infty} \frac{\Delta_i L}{\xi^\text{D}_i}, \quad (V<V_c).
	\label{eq:vFestiL}
\end{align}
Here, we obtain  $v_\text{F}(V)$  by measuring the energy gap $\Delta_{n_f=\pm1}(\mathbf{k}_\text{min})$ of the single-fermion excitation $n_f=1$ at the momentum $\mathbf{k}_\text{min}$ closest to the Dirac point (family H), which shows particularly small finite-size and crossover effects. This momentum corresponds to $\kappa=1$, and the corresponding value of $\xi^\text{D}_i$ is exactly known [see \figref{fig:spec_vs_kappa}(a)]. Note, that  Eq.~(\ref{eq:vFestiL}) cannot be readily applied at $V=V_c$ to extract the critical Fermi velocity $v_\text{F}^c$, since the values of $\xi^\text{CI}_i$ are not a priori known, and cannot be measured independently of $v_\text{F}^c$ in numerical simulations. 
 
\begin{figure}[t]
	\centering
	\includegraphics[width=\columnwidth]{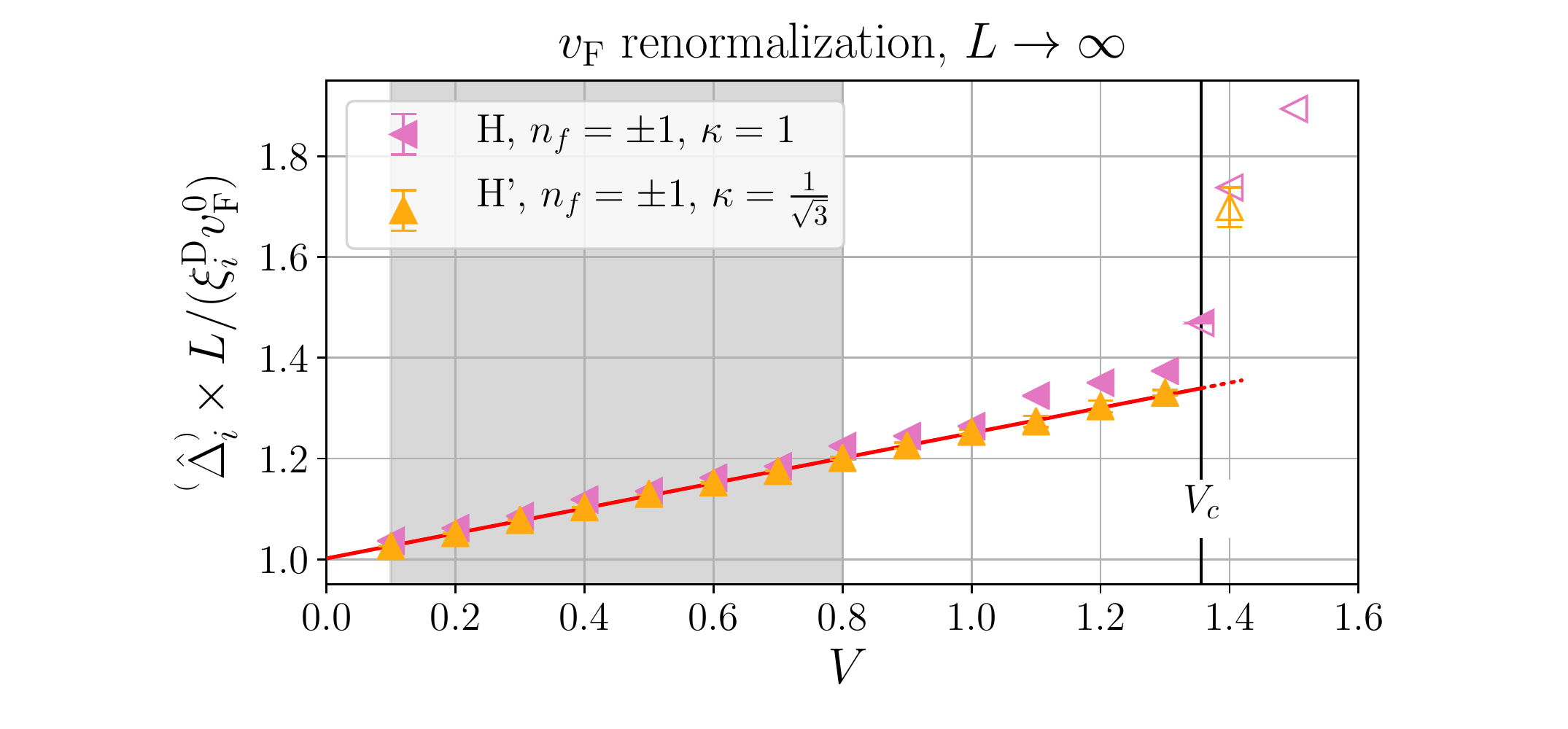}
	\caption{Renormalized Fermi velocity $v_\text{F}(V)$ of the honeycomb lattice model $H_{\rm h}$, based on the energy gap to the single-particle excitation $n_f=\pm1$ at the momentum closest to the Dirac points $\kappa=1$, as obtained after an extrapolation to the TDL. The red line indicates a linear regression of the data points within the shaded region and is an estimator for $v_\text{F}(V)/v_\text{F}^0$ within the SM phase. Based on the data we assume a continous behaviour of $v_\text{F}(V)$ up to the critical point $V_c$ to estimate a critical velocity $v_\text{F}^c$ which is approximately $35\%$ larger than $v_\text{F}^0$. The finite-size results which give the data points in this plot after finite-size extrapolation are shown in \figref{fig:hc_vf_ren}.}
	\label{vfplot}
\end{figure}

The results obtained by the above analysis for the renormalized Fermi velocity $v_\text{F}(V)$ in the SM regime of the honeycomb lattice model $H_{\rm h}$ is shown in Fig.~\ref{vfplot}, along with a linear regression, which describes the numerical results remarkably well. Assuming a non-singular behaviour of the Fermi velocity across the critical point \cite{Herbut2009}, we extrapolate the linear functions to $V_c$ to obtain an estimate for the critical Fermi velocity $v_\text{F}^c$ which is approximately $35\%$ larger than $v_\text{F}^0$. Note that in the GNY field theory the speed of light is the analogue of the Fermi velocity, and stays constant due to strict Lorentz invariance throughout all the phases. 

Furthermore, we perform the same analysis for a gapped level with $n_f=\pm 1$ on clusters in family H', which yields a velocity renormalization in very good agreement with the previous estimate [see \figref{vfplot}].

Alternatively, one may be tempted to consider the slope of the dispersion in the vicinity of the Dirac point, 
\begin{align}\
	v^{\text{slope}}_\text{F}(V) = \lim_{L \rightarrow \infty} \frac{\Delta(\mathbf{k}_\mathrm{min}) - \Delta (\mathbf{K})}{|\mathbf{k}_\mathrm{min}-\mathbf{K}|},
	\label{eq:vf_slope}
\end{align}
to extract the Fermi velocity renormalization: In fact, based on Eq. (\ref{eq:gap_dirac}), we obtain  $v^{\text{slope}}_\text{F}$ in terms of the universal numbers $\xi^\text{D}_i$ for the $n_f=1$ excitations at $\kappa=0$ and $\kappa=1$, respectively. These can be readily extracted from \figref{fig:spec_vs_kappa}(a), and we obtain $v^{\text{slope}}_\text{F}(V) = v_\text{F}(V)$ for $V < V_c $. This estimator is, however, strongly influenced by the crossover effects, mentioned in Sec.~\ref{sec:cross}, close to the QCP: As seen from \figref{fig:spec_vs_kappa}, the single-particle excitation at the Dirac point (with $n_f=1$, $\kappa=0$), which enters the estimator for the Fermi velocity renormalization in  Eq.~(\ref{eq:vf_slope}), shows particularly  different values of $\xi^\text{D}_i$ and $\xi^\text{CI}_i$. Hence, performing numerical simulations on insufficiently large tori (due to limitations in the accessible system sizes), will lead to a false estimate for the Fermi velocity renormalization by Eq.~(\ref{eq:vf_slope}).

An extrapolation of the Fermi velocity based on slopes between the Dirac point and the closest momentum nearby, as in Eq.~(\ref{eq:vf_slope}), is thus particularly dangerous in the vicinity of interacting quantum critical points and a very careful analysis including proper finite-size scaling is necessary~\cite{Sen2015, Schuler2016}. As pointed out recently~\cite{Hesselmann2019}, such a crossover due to enhanced finite-size shifts in the excitation gap at the Dirac point led the authors in Ref.~\cite{Tang2018} to drastically underestimate the Fermi velocity in the Hubbard model on the honeycomb lattice near its QCP. While the nature of the quantum critical point is different in Ref.~\cite{Tang2018} (chiral Heisenberg fixed point), the problem of the crossover length scale also applies there.

\section{Numerical Results for the Lattice Models}
\label{sec:hc_numerics}

In this section, we present our numerical results for the microscopic lattice models in more detail. We begin by providing an overview of the two  methods that we used for our numerical calculations, exact diagonalization (ED) and quantum Monte Carlo (QMC), and explain how we corrected for  warping effects from the lattice discretization, before providing details of the extrapolation to the thermodynamic limit.

\subsection{Exact Diagonalization}
\label{sec:hc_numerics_ed}

Exact diagonalization (ED) \cite{Lauchli2011,Wietek2018} can be used to calculate all low-energy gaps directly and exactly for all parameters $V$ on finite size clusters. In addition, their quantum numbers according to fermion-number conservation and lattice symmetries can be directly identified using a symmetry-adapted basis. In particular, the ED spectrum is divided into $n_f$ sectors combined with a $\mathbb{Z}_2$ charge according to the particle-hole symmetry $C$ at half filling $n_f=0$,  as well as the momentum quantum number $\mathbf{k}$, and an irreducible representation of the lattice' point-group. This also allows for the identification of appropriate quantum many-body operators with non-vanishing matrix elements between the ground state and the various low-lying excitations. These operators can then be used to extract the corresponding energy gaps within the QMC simulations from the decay of the imaginary-time correlation function, as described in the next section.

\subsection{Quantum Monte Carlo}
\label{sec:hc_numerics_qmc}

We employ the projector lattice continuous-time quantum Monte Carlo algorithm (LCT-INT) detailed in~\cite{Wang2015}. The ground state expectation value of an observable $\hat{O}$ is accessed upon projecting a trial wave function $| \Psi_T \rangle$,
\begin{align}
	\frac{\left\langle \Psi_0 \right| \hat{O} \left| \Psi_0 \right \rangle}{\left \langle \Psi_0 | \Psi_0 \right \rangle} =
        \lim_{\Theta \rightarrow \infty} \frac{\left \langle \Psi_T \right| \E^{-\frac{\Theta}{2} H} \hat{O}\, \E^{-\frac{\Theta}{2} H} \left| \Psi_T \right \rangle}{\left \langle \Psi_T \right| \E^{-\Theta H} \left|\Psi_T \right \rangle} \ ,
\end{align}
where $|\Psi_0\rangle$ denotes the ground state of the Hamiltonian $H$. For this work, the simulations were performed with a projection length of up to $\Theta = 160 / t$ to ensure convergence within the statistical uncertainty. Importantly, the LCT-INT formulation does not rely on a Trotter decomposition, but instead decomposes the projection operator using an interaction expansion directly in continuous time, thus eliminating the Trotter error completely. 

The trial wave function is chosen as a zero momentum, particle-hole (anti-)symmetric ground state of the free Hamiltonian, and is represented by a Slater determinant $| \Psi_T \rangle$. Furthermore, the invariance of the Hamiltonian under the reflection symmetry $I_x$  can be used to separate the two quasi-degenerate ground states of the interacting system. We therefore consider trial wave functions $| \Psi_T \rangle_{\pm}$ with $I_x$-eigenvalue $\pm 1$ in order to project onto the ground state of the corresponding symmetry sectors.

The lowest energy gaps $\Delta_{\hat{O}}$ are extracted from the asymptotic decay of imaginary-time correlation functions, dominated by 
\begin{align}
	\left \langle \hat{O}(\tau) \hat{O}^\dag \right \rangle \sim \left| \left\langle \Psi_0 \right| \hat{O} \left| \Psi_{\Delta_{\hat{O}}} \right\rangle\right|^2 \exp(- \Delta_{\hat{O}} \tau) \label{eq:OOdag}\,,
\end{align}
where for sufficiently large $\tau$, with $1 \ll \tau \ll \Theta/2$, this leading exponential decay is dominated by the smallest gap $\Delta_{\hat{O}}$ accessible by the operator $\hat{O}$. The relevant energy gaps correspond to excited states in different symmetry sectors, which are determined by their fermion-number, momentum, particle-hole symmetry, and an irreducible representation of the lattice point-group. The operators $\hat{O}$ have to connect the ground state $| \Psi_{0} \rangle$ to the desired excited state $| \Psi_{\Delta_{\hat{O}}} \rangle$  such that the overlap $| \langle \Psi_0 | \hat{O} | \Psi_{\Delta_{\hat{O}}} \rangle|$ is finite. Feasible operators $\hat{O}$ can be categorized by their action under the various symmetry operations. For example, states with opposite parity under $I_x$  are connected by operators for which the anti-commutator with the reflection operator $I_x$ vanishes, $\{ \hat{O}, I_x \} = 0$. A detailed list of the symmetry properties of the operators connecting the ground state to the various relevant excited states can be found in Table~\ref{tab:sym}. The explicit expressions of possible operator implementations can be found in Appendix~\ref{app:qmc_operators}.

\subsection{Warping Corrections}
\label{sec:hc_numerics_warping}

\begin{figure}[htb]
	\centering
	\includegraphics[width=\columnwidth]{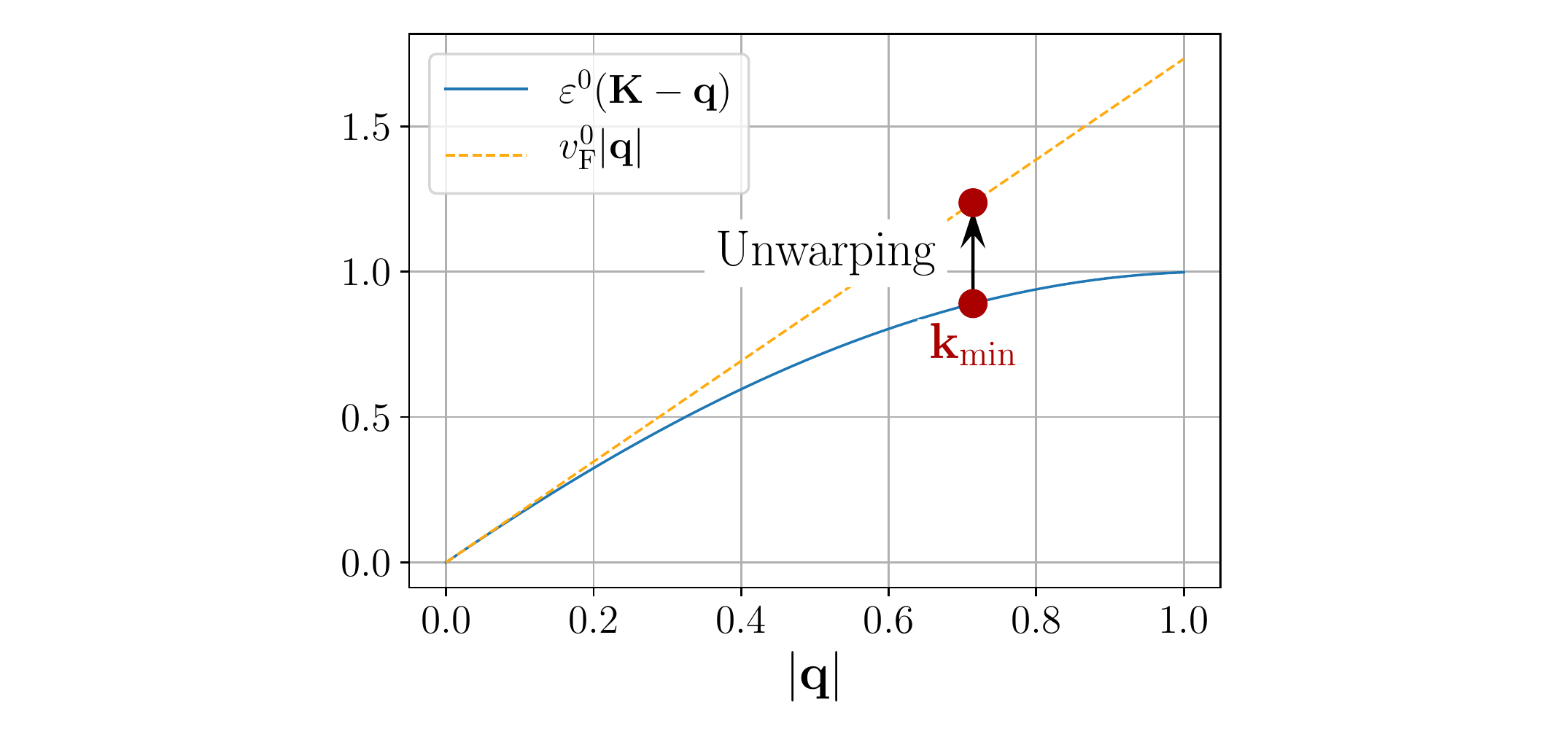}
	\caption{Unwarping of the energy spectrum for clusters that do not contain the Dirac points in order to reduce finite-size effects. The finite-size spectrum for all interaction strengths $V$ is multiplied by a ($\ns$-dependent) constant such that the non-interacting fermion dispersion aligns with the (effective) linear Dirac cone with velocity $v_\text{F}^0$ at $\mathbf{k}=\mathbf{k}_\text{min}$. See the main text for further details.}
	\label{fig:disp_ren}
\end{figure}

It is well known that the Dirac cones for the non-interacting lattice models, $V=0$, are strongly modified by warping effects from the asymptotic dispersion relations, see~\figref{fig:gnymodels}. As a result, rather large systems are required to directly probe the linear dispersion regime, especially for those clusters that do not contain the Dirac points. In order to reduce the finite-size effects for the energy levels measured on clusters without Dirac points, we appropriately rescale the finite-energy spectra at any momentum $\mathbf{k}$  in the vicinity of the Dirac points for all interaction strengths $V$ as
\begin{equation}
    \Delta_i \rightarrow \hat\Delta_i=\frac{v_\text{F}^0 \left|\mathbf{D} - \mathbf{k}_\text{min} \right|}{\varepsilon^0\left(\mathbf{k}_\text{min}\right)} \Delta_i\,, 
\end{equation}
where $v_\text{F}^0$ denotes the Fermi velocity for $V=0$,  $\varepsilon^0(\mathbf{k})$  the dispersion relation of the non-interacting system, and $\mathbf{D}$ is equal to $\mathbf{K}$ or $\mathbf{X}$, for $H_{\rm h}$ and $H_{\rm s}$, respectively. This unwarping of the energy spectrum is illustrated in \figref{fig:disp_ren}. In the following, $\hat\Delta_i$ will always indicate that an unwarping according to the above equation has been performed. 

\subsection{Results - Evolution of the Torus Spectrum of $H_{\rm h}$ with $V$}
\label{sec:hc_numerics_evol}

In this section, we provide a detailed analysis of the spinless fermion $t-V$ model on the honeycomb lattice, described by \eqref{eq:hcmodel} at half-filling. The underlying structure of the energy level spectrum  is uncovered upon
appropriately rescaling the finite-size energy gaps by a factor of the linear system scale, as quantified by $L=\sqrt{N_s/2}$.
We, therefore, consider in the following the low-energy gaps rescaled as $\Delta_i \times L$, which we  call the spectrum.
We first consider the evolution of the full low-energy spectrum with the interaction strength $V$, including the free system $V=0$ and the 
QCP at $V_c \approx 1.355$~\cite{Wang2014, Li2015}, based on ED calculations on clusters of a few ten sites.
This is very instructive in order to identify the qualitative features of a QCP that remain valid in the thermodynamic limit, even though the quantitative values of the energy gaps may be subject to substantial finite-size effects. 

\begin{figure*}[htb]
	\centering
	\includegraphics[width=\textwidth]{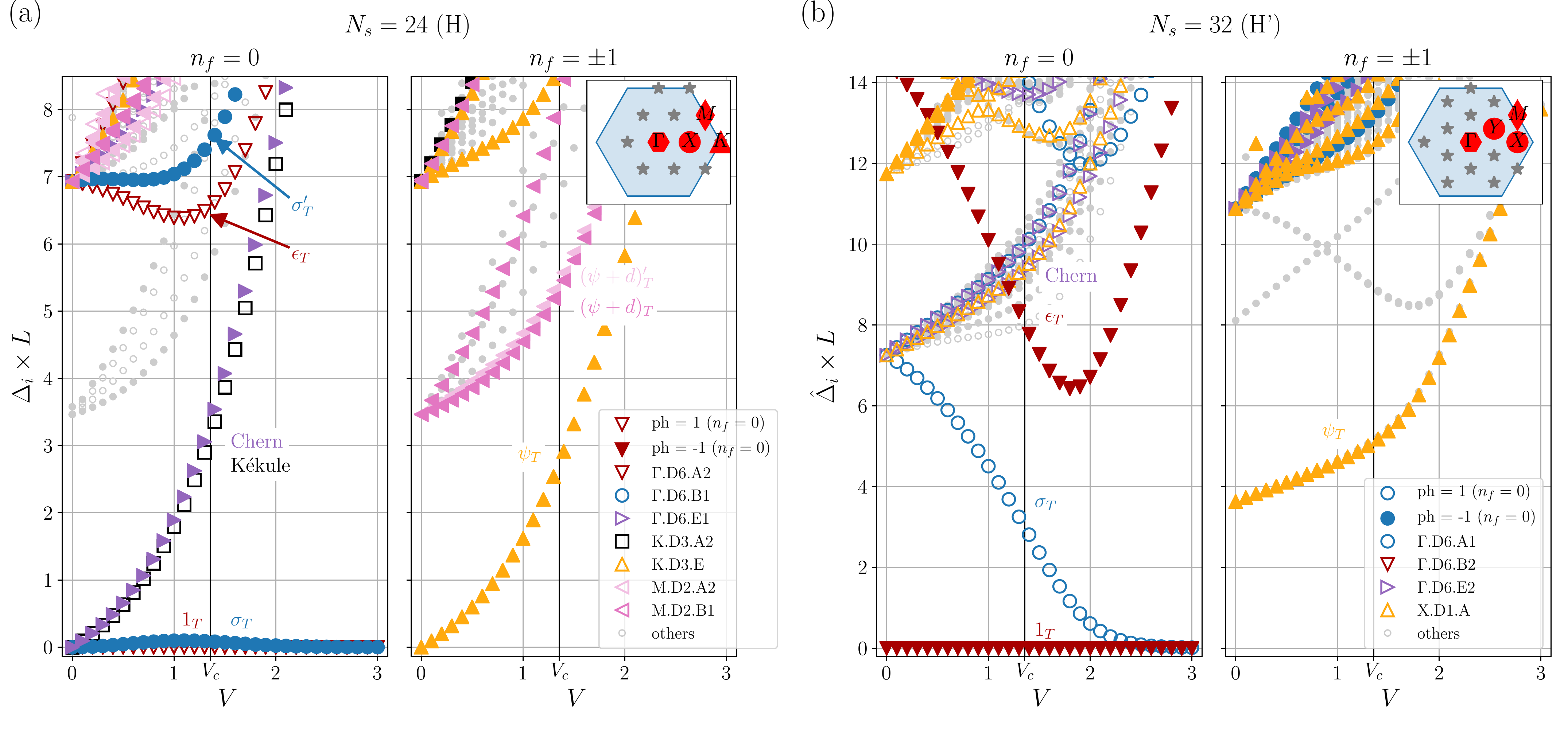}
	\caption{Low-energy spectra of the model \eqref{eq:hcmodel} as a function of $V$ for clusters of size (a) $N_s=24$, (b) $N_s=32$. The left panels show the half-filled sector $n_f=0$, the right panels show the sectors with one additional fermion/hole $n_f=\pm1$. The black vertical line indicates the critical point $V_c \approx 1.355$~\cite{Wang2014, Li2015}. The cluster in (a) has the Dirac point in its momentum space (family H) and the fermionic excitation (right panel) is gapless for $V=0$.  The cluster in (b) does not feature the Dirac point (family H') and the fermionic excitation is always gapped. This influences the spectrum at criticality such that different families of clusters have to be distinguished in the extrapolation to the TDL. We have also, in the corresponding colors, indicated the labels for the most important low-energy spectral levels: $1$, $\sigma_T$, $\epsilon_T$, $\psi_T$, Chern, Kekul\'e (see text). Empty (filled) symbols for $n_f=0$ represent even (odd) levels under particle-hole inversion.}
	\label{fig:hc_fullspecs}
\end{figure*}

\figref{fig:hc_fullspecs}(a) shows the evolution of the  spectrum on the ${N_s=24}$ sites cluster, which is characteristic of tori   that contain the Dirac points (family H). In the non-interacting case, ${V=0}$, we identify a six-fold degenerate ground state in the half-filled sector, ${n_f=0}$. The lowest single-fermion level ${n_f=\pm1}$~\footnote{Due to particle-hole symmetry the spectrum is identical for ${n_f=\pm \nu, \, \nu \in \mathbb{Z}_0}$, i.e., if we add $\nu$ fermions or holes. Therefore, we only show the $n_f\geq0$ sectors.} (8-fold degenerate) and the two-fermion sector ${n_f=\pm2}$ (two-fold degenerate; not shown) are also gapless, as fermions can be created at the gapless Dirac points. The finite-size spectrum immediately gaps out for finite ${V>0}$, and the system undergoes a transition into the CDW phase for ${V>V_c}$, where a two-fold (quasi-)degenerate ground state of a $\mathbb{Z}_2$ even and a $\mathbb{Z}_2$ odd level is observed, while all fermionic excitations ${n_f\neq0}$ are gapped.

According to their quantum numbers we can label the energy levels and relate them to the known instabilities of the Dirac phase: Mass gaps for spinless Dirac fermions can be generated by breaking the sublattice symmetry ($\sigma_T$), by breaking the time-reversal symmetry with zero net magnetic flux through the honeycomb unit cell (Chern), or by a Kekul\'e dimerization which creates two distinct real masses~\cite{Semenoff1984,Haldane1988,Hou2007,Ryu2009}. 

Another prominent level near the quantum critical point is the state corresponding to a detuning from the quantum critical point ($\epsilon_T$). This level lies in the same symmetry-sector as the ground state ($1_T$) and typically shows a characteristic shape with a minimum around the QCP. This level is also the leading contribution to the fidelity susceptibility at the quantum critical point~\cite{Albuquerque2010,Wang2015b}.

The critical torus spectrum in the past showed qualitatively similar structures as the operator content of the corresponding field theory, i.e., the scaling dimensions of the fields of the GNY CFT~\cite{Schuler2016,Whitsitt2017}. We have, therefore, chosen the labels $\sigma_T$ ($\epsilon_T$) as the torus analogues of the lowest particle-hole odd (even) scalar fields and $\psi_T$ as the torus analogue of the lowest vector field (the lowest single-fermion excitation ${n_f=\pm1}$). Furthermore, we label the lowest two-fermion excitation ${n_f=\pm2}$ as $(2\psi)_T$ and the torus analogue of the fermionic descendent field, i.e., the lowest fermionic excitation at $\mathbf{k}_\text{min}$, as $(\psi+d)_T$. A prime  on a level symbol is used to indicate the second-lowest level in the same symmetry sectors as the corresponding unprimed level. In Tab.~\ref{tab:gny_levels} we list the most important quantum numbers for these levels.

\begin{table}[t]
    \centering
    \begin{tabular}{|c|c|c|c|c|c|}
    \hline
        Levels & PH  & $I_x$ & $n_f$ & $\mathbf{k}$ & $\quad \kappa \quad $\\
    \hline \hline
        $1_T,\epsilon_T$ & $1$ & $1$ & $0$  & $0$ & $0$ \\
    \hline
    $\sigma_T,\sigma_T^\prime$ & $-1$ & $-1$ & $0$  & $0$ & $0$ \\
    \hline
     Kekule & $1$ & $1$ & $0$ & $\mathbf{K}$ & $0$ \\
    \hline
    Chern & $-1$ & $-1$ & $0$  & $0$ & $0$ \\
    \hline
    $\psi_T$ & -- & -- & $\pm 1$ & $\mathbf{K}$ & $0$ \\
    \hline
    $(\psi+d)_T,(\psi+d)_T^\prime$ & -- & -- & $\pm 1$  & $\mathbf{k}_\mathrm{min}$ & $1$ \\
    \hline
    $(2\psi)_T$ & -- & -- & $\pm 2$ & $0$ & $0$ \\
    \hline
    \end{tabular}
    \caption{Quantum numbers  of the most
        relevant energy levels in the torus spectrum of the chiral Ising CFT. The table denotes the particle-hole quantum
        number PH, the fermion sector relative to half filling $n_f$, the momentum $\mathbf{k}$, and the reduced momentum
        $\kappa$. For simplicity, we here omit showing the irreducible representations under the lattice point-group
        symmetry.}
    \label{tab:gny_levels} 
\end{table}

The finite-size spectrum for the family of clusters that do not include the Dirac points is structurally different in the SM phase [see \figref{fig:hc_fullspecs}(b)]. The ground state is unique even for ${V=0}$ and the fermion excitations ${n_f=\pm1}$ and ${n_f=\pm2}$ are gapped. The $\sigma_T$ field strongly decreases in energy with a finite value at the critical point and constitutes the second state in the two-fold degenerate ground state manifold for ${V \gg V_c}$, but, in contrast to the clusters with Dirac points has a large gap at $V_c$. Again, the $\epsilon_T$ field shows a very characteristic shape with a strongly reduced gap only around the critical point.

\subsection{Results -  Torus Spectrum at Criticality}
\label{sec:hc_numerics_spec}

\begin{figure}[ht]
	\centering
	\includegraphics[width=\columnwidth]{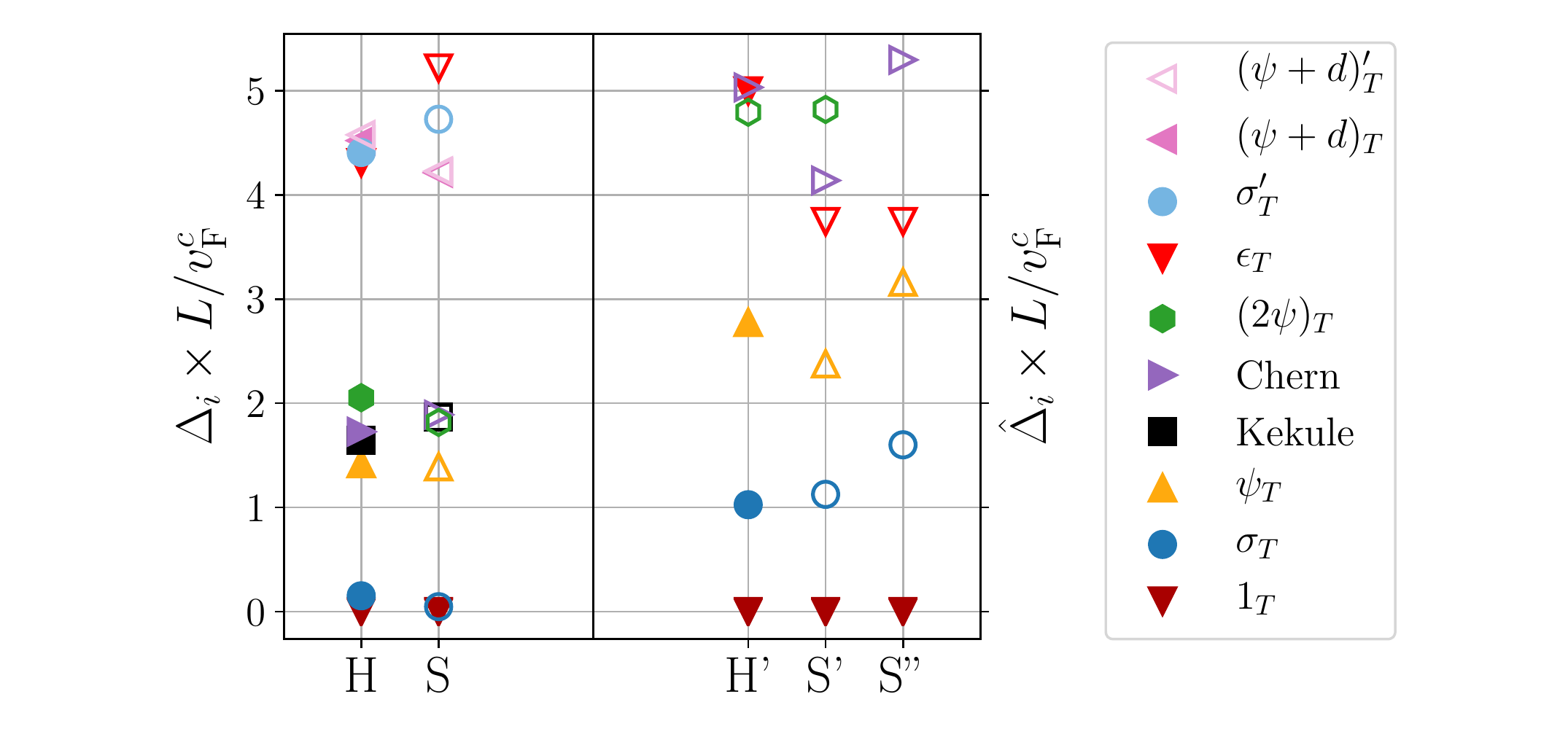}
	\caption{Critical torus spectrum of the $N=4$ chiral Ising theory for the honeycomb and square models. Shown are the most prominent low-energy levels after finite-size extrapolation [c.f. \figref{fig:hc_extrap}, \figref{fig:sq_extrap}]. The left panel  shows the extrapolated levels from tori that contain the Dirac points, while the right panel shows those that do not contain the Dirac points. The spectrum has been normalized by the critical Fermi velocities $v^c_\text{F}$ for the different geometries [c.f. \figref{fig:hc_vf_ren}, \figref{fig:sq_vf_ren}]. Full symbols show levels extrapolated from QMC data, empty symbols denote levels extrapolated from ED data alone.}
	\label{fig:overview}
\end{figure}

We next examine in more detail the spectrum at criticality for the two microscopic lattice models, $H_{\rm h}$ and $H_{\rm s}$. In \figref{fig:overview} we show the critical torus energy spectra, as obtained from the different cluster geometries. In this figure, the critical torus energy spectra are rescaled by the critical Fermi velocity, which, up to a global factor, identifies the universal numbers $\xi_i^\text{GNY}$ for the $N=4$ chiral Ising GNY universality class in $D=(2+1)$ dimensions. As mentioned in Sec.~\ref{sec:torus}, one  furthermore has to distinguish for each model different families of finite size clusters. For the honeycomb lattice, the families H (H')  are distinguish by the presence (absence) of the Dirac points in their momentum space [c.f. \figref{fig:lattice_families}(a)]. For the square lattice, we distinguish three families:  Clusters in family S contain the two Dirac points among the lattice momenta, while for those belonging to family S' (S''), the Dirac points are (not) located at the center between four lattice momenta, respectively [c.f. \figref{fig:lattice_families}(b)]. 

\begin{figure*}[htb]
	\centering
	\includegraphics[width=\textwidth]{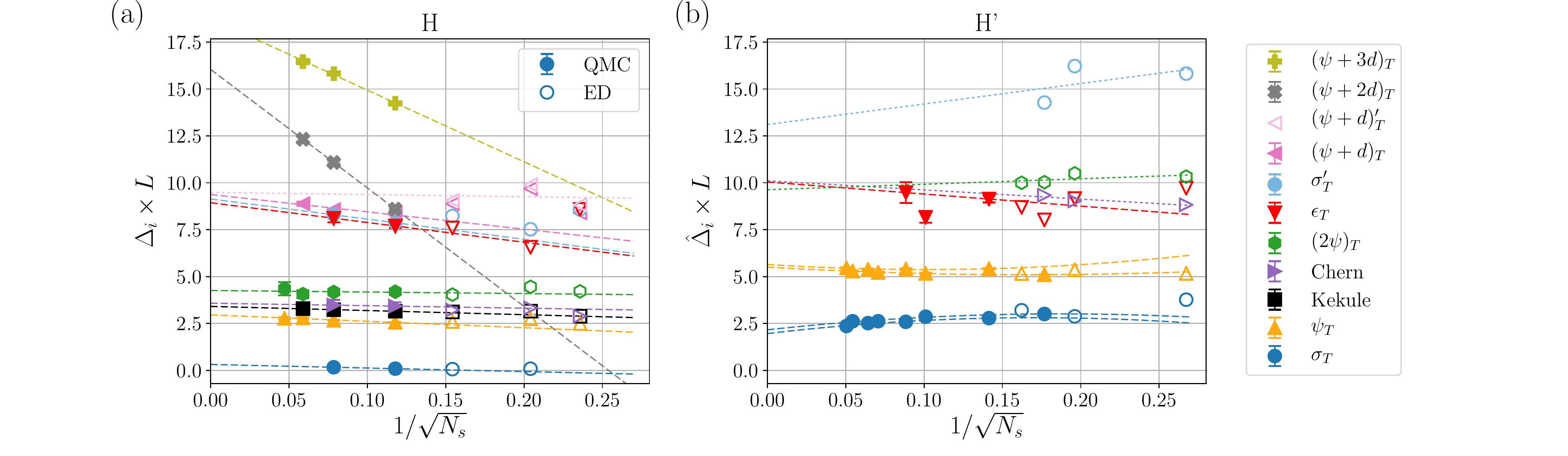}
	\caption{Extrapolation of the critical energy spectrum on the honeycomb lattice. We use second or first order polynomial functions (depending on the number of available data points) in $1/\sqrt{N_s}$ to extrapolate the gaps to  $N_s\rightarrow \infty$. Open symbols denote ED data, while full symbols show QMC data, and we use the same labelling for the levels as in \figref{fig:hc_fullspecs}. Dashed (dotted) lines show extrapolations of QMC (ED) data. (a) shows clusters that contain the Dirac points, (b) shows clusters that do not. In (b) we extrapolate $L\mod3=1$ and $L\mod3=2$ clusters separately, where possible. We use clusters of linear size up to (a) $L=15$, (b) $L=14$ for the extrapolation. Note, that not all gaps could be obtained on the largest clusters with QMC due to small overlap of some excitations with the ground state for the chosen operators, which resulted in noisy estimators and prevented us to reliably extract gaps for the largest systems. } 
	\label{fig:hc_extrap}
\end{figure*}

To obtain the critical torus spectrum in the thermodynamic limit we extrapolated the finite-size results $\Delta_i \times L $ for the different levels [c.f. also \figref{fig:hc_fullspecs}] obtained from ED and QMC to $N_s\rightarrow\infty$, as shown in \figref{fig:hc_extrap} for the model $H_{\rm h}$, on which we focus in the remainder of this section. The details of the corresponding finite-size analysis of the torus spectrum at the QCP of the square lattice model $H_{\rm s}$ are provided in Appendix~\ref{app:sq_numerics}. While the quality of the extrapolations is not equally good for all levels since not all energy gaps could be measured with QMC, it is important to note that the qualitative structure of the low-energy levels, their quantum numbers and (approximate) multiplicities are already present on the smaller clusters with a few tens of sites, and it is this qualitative structure that serves as a fingerprint for the chiral Ising universality class. For the same reason we also omit to give error bars on the extrapolated levels. 

The critical torus energy spectrum for tori  that contain the Dirac points (family H) show very characteristic features [c.f. Figs.~\ref{fig:overview},~\ref{fig:hc_extrap}]: The gap for the $\sigma_T$ field is remarkably small and appears to form a two-fold degenerate state together with the ground state, i.e., the vacuum level $1_T$ in the thermodynamic limit. The $\epsilon_T$ level and the $\sigma_T^\prime$ level are also close to each other and build a second copy of such a two-fold (nearly) degenerate level in the thermodynamic limit. The fermion mode $\psi_T$ is the next lowest level above $\sigma_T$. The subsequent Kekule, Chern and $(2\psi)_T$ levels are very close to each other and build a six-fold (nearly) degenerate level. Two single-fermion levels $(\psi+d)_T^{(\prime)}$ with momentum $\mathbf{k}_\text{min}$ are also found to build a nearly degenerate set of levels with energy comparable to the $\epsilon_T$ and $\sigma_T^\prime$ states. The characteristic two-fold nearly degenerate levels are a result of the two-fold nearly degenerate ground state.

For  tori  that do not contain the Dirac points (family H'), the critical torus energy spectrum is strongly altered, with a much larger $\sigma_T$ gap, followed by the $\psi_T$ field. The opening of the $\sigma_T$ gap also leads to a strong splitting of the other two-fold nearly degenerate levels observed on the clusters with Dirac points. The $(2\psi)_T$ and Chern fields are, again, very close to each other, while the Kekule field is not clearly defined. The $\epsilon_T$ level seems to be only slightly influenced by the choice of the torus  shape.


\begin{figure*}[t]
 \centering
 \includegraphics[width=\textwidth]{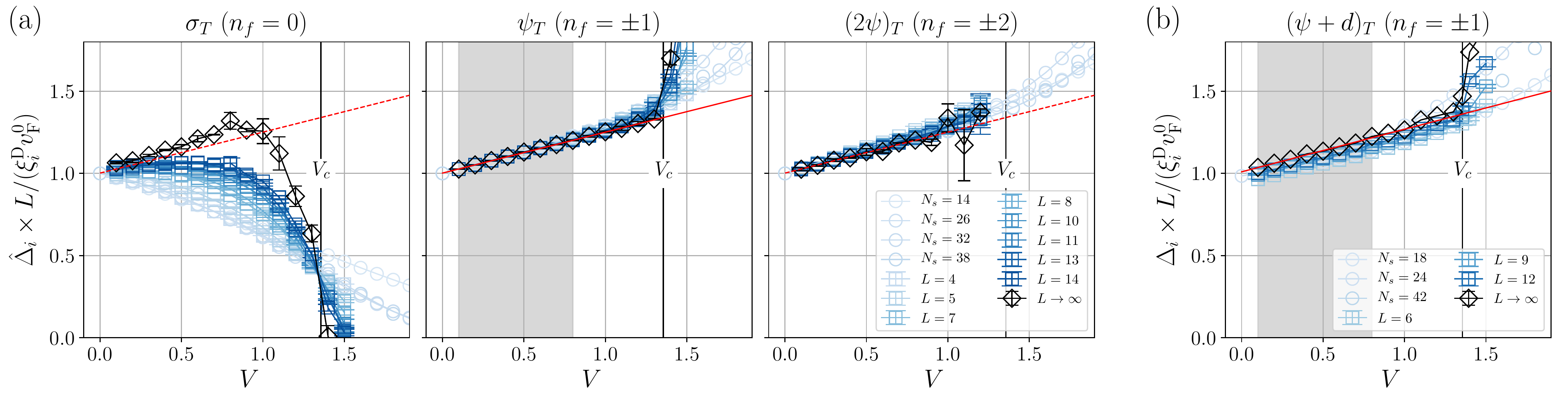}
 \caption{Renormalization of the Fermi velocity in the SM phase for the honeycomb lattice model. 
     (a) Energy
     gaps $\hat{\Delta}_i \times L$ of low-lying levels for clusters without Dirac points (family H') as a function of the interaction strength $V$. 
     The gaps are normalized to one at $V=0$ for easier comparison.
     Blue square (circle) symbols show finite-size data from QMC (ED), while the black diamonds are the finite-size
     extrapolated values ($L\rightarrow \infty$) for each $V$. The extrapolated data for the fermion excitation $n_f=1$ (center panel) is
     fitted by a linear function (full red line), showing the Fermi velocity renormalization.This fit is included in
     the other panels by a dashed line as a comparison.
     The shaded region indicates the fitting window.
     (b) Energy gap of the single fermion excitation at
     $\mathbf{k}_\text{min}$ for clusters  that contain the Dirac points (family H),  together with a fit describing 
     the $v_\text{F}$ renormalization. Note, that the gaps for levels shown in (a) would vanish in the SM phase for these 
     clusters. See text for further discussion.}
 \label{fig:hc_vf_ren}
\end{figure*}

\subsection{Results - Estimation of the Fermi Velocity Renormalization}
\label{sec:hc_numerics_vfren}

We already discussed our approach to estimate the Fermi velocity renormalization and the subtle crossover effects near the QCP in Secs.~\ref{sec:vf_ren} and \ref{sec:cross}. Here, we provide further details on this procedure for $H_{\rm h}$, while the case of $H_{\rm s}$ is treated in Appendix~\ref{app:sq_numerics}. In particular, we consider using different levels to access $v_\text{F}(V)$ based on the general formula \eqref{eq:vFestiL} within the SM phase. Here, we focus on the Hamiltonian $H_{\rm h}$, for which larger-system QMC data is available. 

In \figref{fig:hc_vf_ren}(a) we show various renormalized energy levels $\Delta_i\times L$ within the SM phase of the honeycomb lattice model $H_{\rm h}$ from finite clusters that do not contain the Dirac points, and also extrapolated these values to the thermodynamic limit $N_s\rightarrow \infty$.  Because of the very small finite-size effects of the $n_f=\pm1$ level ($\psi_T$), the Fermi velocity renormalization is best read off from this level and its $V$-dependence can be well approximated by a linear function [see center panel in \figref{fig:hc_vf_ren}(a)].  Furthermore,  the renormalization of other levels with stronger finite-size effects [see left and right panels in \figref{fig:hc_vf_ren}(a)] is well approximated by the same function in the SM phase. It is important to note, that the strong drop of the $\sigma_T$ level close to the critical point is not due to a sudden, strong decrease of the Fermi velocity, but mainly because of the strong difference of the universal numbers $\xi^\text{D}_{\sigma_T}$ in the SM phase and $\xi^\text{GNY}_{\sigma_T}$ at the chiral Ising critical point. This level, thus, provides us with another dramatic example of the crossover effects that have been discussed in Sec.~\ref{sec:cross}.

For clusters that contain the Dirac points, many gaps, such as the single particle gap at the Dirac point ($\psi_T$), vanish faster than $1/L$ in the SM phase, i.e., these levels have a vanishing value of $\xi^\text{D}_i=0$, and they, thus, cannot be used to extract $v_\text{F}(V)$ based on \eqref{eq:vFestiL}. On these clusters, we, therefore, measure the gap of a single fermion excitation $n_f=1$ with $\kappa=1$, labeled $(\psi+d)_T$, which is gapped, and obtain the $v_\text{F}(V)$ renormalization from this value [see \figref{fig:hc_vf_ren}(b)]. The extrapolated values and the linear regression function is also shown in Fig.~\ref{vfplot} and was further discussed in Sec.~\ref{sec:vf_ren}.  Again, we obtain an approximately linear behaviour and, importantly, the regression functions agree very well among the two families of finite-size clusters. 
The agreement of $v_\text{F}(V)$ as extracted using  different levels and cluster families demonstrates that this approach of computing the Fermi velocity renormalization within the SM phase is quite reliable.

\section{Torus Spectrum in the $\epsilon$ Expansion}
\label{sec:eps_expansion}

In this section, we provide the details of the analytical $\epsilon$ expansion to extract the torus spectrum for the GNY field theory. 

\subsection{General Structure of the Expansion}
\label{sec:eps_expansion_structure}

We want to examine the finite-size spectrum of the GNY field theory given in \eqref{eq:LGNY}. For this purpose, we will use a real-time Hamiltonian formulation, with
\bea
	H_\text{GNY} &=& \int \dd^{d} x \, \Bigg[ \overline{\Psi}^j  \left( \cancel{\nabla} + g_{\rm Y} \, \phi^j \right) \Psi \nn
	&& \qquad \qquad + \ \frac{1}{2} \Pi^2 + \frac{1}{2} \left( \nabla \phi \right)^2 + \frac{\lambda}{4!} \phi^4 \Bigg].
	\label{eq:HGNY}
\eea
These operators satisfy the equal-time commutation relations
\bea
	\left[\phi(x),\Pi(x')\right] &=& \I \delta^d(x - x'), \nn
	\left\{ \Psi^j_a(x),\Psi_b^{j \, \dagger}(x') \right\} &=& \delta^{j j'}\delta_{ab} \delta^d(x - x'),
\eea
where the Dirac field $\Psi_a^{j}$ has spinor index $a = 1, ... ,n_\text{D}$ and flavor index $j = 1,...,N_f$, and we define $N = n_\text{D} N_f$ to be the total number of degrees of freedom. In \eqref{eq:HGNY} we have assumed that the tuning parameter $s$ has already been set to its critical value $s_c$, and used the fact that $s_c=0$ in dimensional regularization. At leading order in $\epsilon = 3 - d$, the interaction couplings flow to the fixed point values~\cite{zinn1991}
\bea
	g_{\mathrm{Y}, \mathrm{ren.}}^{\ast 2} &=& \frac{16 \, \pi^2 \epsilon}{N+6}, \label{eq:gnyfixedpoint}\\
	\lambda^{\ast}_\text{ren} &=& \frac{ 384 \, N \pi^2 \epsilon }{(N+6)\left[ (N - 6) + \sqrt{N^2 + 132 N + 36} \right]}.\nonumber
\eea

As a reminder of our notation, we use complex coordinates for the two-dimensional torus, $x = x_1 + i x_2$, and then define the torus by two complex periods, $\omega_1$ and $\omega_2$, such that the points $x + n \omega_1 + m \omega_2$ are equivalent for all $n,m \in \mathbb{Z}$. The torus is characterized by its modular parameter, $\tau \equiv \omega_2/\omega_1 = \tau_1 + i \tau_2$, and its area is given by $\mathcal{A} = \mathrm{Im}\left( \omega_2^{\ast} \omega_1 \right)$. In \eqref{eq:HGNY}, we take the spatial integral to be over $d/2$ copies of the two-dimensional torus with modular parameter $\tau$, which preserves point group symmetries at all steps of the calculation, and does not introduce any extra unphysical parameters. In addition to the other symmetries mentioned in this paper, we note that the full torus spectrum is also invariant under the modular transformations, $\tau \rightarrow \tau + 1$ and $\tau \rightarrow -1/\tau$, under which the torus area $\mathcal{A}$ is also invariant. The length scale $L$ introduced in earlier sections to define the universal numbers $\xi_i^{\text{CI}}$ are related to the area of the torus by $\mathcal{A} = |\mathbf{a}_1| \tau_2 L^2$ (see Figure \ref{fig:lattice}).

As mentioned in Section \ref{sec:torus}, the structure of the $\epsilon$ expansion on the torus turns out to be very different if we allow twisted boundary conditions. In particular, if we consider torus clusters without Dirac points, this corresponds to a boundary condition $\Psi(x + n \omega_1 + m \omega_2) = e^{i \theta} \Psi(x)$ on the fermions, while the bosonic field remains fully periodic. The twisted boundary condition results in a finite-size mass gap for the fermions proportional to $\theta$, so one does not need to separate out the fermionic zero modes. Then following the arguments in Section III.A. of Ref.~\onlinecite{Whitsitt2017}, whenever $\theta^2 \gtrsim \epsilon^{1/3}$, the torus spectrum is given by  an effective Hamiltonian which only involves the bosonic zero modes, implying that the torus clusters which do not have Dirac points will have a dramatically different spectrum from the $L\!\!\mod3=0$ case corresponding to the field theory with periodic boundary conditions. In the remainder of this section, we will only focus on the fully periodic setup.

As emphasized in previous work~\cite{Schuler2016,Whitsitt2017}, the presence of massless bosonic fields invalidates naive perturbation theory in a finite volume. To obtain the finite-size spectrum, one must separate out the zero-momentum part of the fields, and subsequently treat the interactions of these zero-modes \emph{exactly}. The non-perturbative treatment of the zero modes results in a torus spectrum which is dramatically different from the particle-like Fock spectrum of the Dirac case. To this end, we write the mode expansions for our fields as
\bea
	\phi(x) &=& \mathcal{A}^{\frac{1 - d}{4}} \varphi + \frac{1}{\mathcal{A}^{d/4}} \sum_{k \neq 0} \frac{\E^{\I k \cdot x}}{\sqrt{2 |k|}} \left[ a(k) + a^{\dagger}(-k) \right],  \nn
	\Pi(x) &=& \mathcal{A}^{-\frac{d+1}{4}} \pi - \frac{i}{\mathcal{A}^{d/4}} \sum_{k \neq 0} \sqrt{\frac{|k|}{2}} \E^{\I k \cdot x} \left[ a(k) - a^{\dagger}(-k) \right],  \nn
	\Psi^{j}_a(x) &=& \mathcal{A}^{-\frac{d}{4}} \psi^j_a + \frac{1}{\mathcal{A}^{d/4}} \sum_{k \neq 0} \sum_{s = 1}^{n_\text{D}/2} \frac{\E^{\I k \cdot x}}{\sqrt{2 |k|}} \Big[ u_a(k,s) b^{ \, j}_s(k) \nn
	&& \qquad \qquad \qquad \qquad + \ v_a(-k,s) c^{\, j \dagger}_s(-k) \Big].
\eea
Here, $a^{\dagger}$, $b_s^{\, j \dagger}$, and $c_s^{\, j \dagger}$ create Fock states for bosons, fermions, and anti-fermions respectively (and the fermions have pseudospin and flavor indices $s$ and $j$). Dot products for complex coordinates are defined as $k \cdot x \equiv \mathrm{Re}\left( k x^{\ast} \right)$. The momentum sums are performed over the reciprocal lattice, which is given by
\beq
	k = n k_1 + m k_2, \qquad n,m \in \mathbb{Z},
\eeq
where $k_1 = -i\omega_2/\mathcal{A}$, $k_1 = i\omega_1/\mathcal{A}$.
The commutation relations for the zero-mode parts are
\bea
	[\varphi,\pi] = \I\,, \qquad  \{ \psi^j_a,\psi_b^{\, j' \dagger} \} = \delta^{j j'}\delta_{ab}.
\eea

We now place the mode expansion into \eqref{eq:HGNY} and separate the Hamiltonian into an unperturbed and interacting part, $H_\text{GNY} = H_0 + V$, where we insist that all zero-mode operators are included in $V$. Then $H_0$ is a free Fock Hamiltonian,
\bea
	H_0 &=& E_0 + \sum_{k \neq 0} |k| \, a^{\dagger}(k) a(k) \nn
	&& + \ \sum_{k \neq 0}\sum_{s = 1}^{n_\text{D}/2} |k| \left[ b_s^{\, j \dagger}(k) b^j_s(k) + \ c_s^{\, j \dagger}(k) c^j_s(k) \right],
	\label{eq:fockham}
\eea
where $E_0 = -(3/2) \sum_{k \neq 0} |k|$ is the leading contribution to the ground state energy, and repeated flavor indices are always summed from $j = 1, ..., N_f$. It is possible to compute the universal part of the ground state energy (the calculation for the Wilson-Fisher CFT is given in Ref.~\onlinecite{Whitsitt2017}), but in this paper we will only compute the energy splittings from the ground state, and hereafter we subtract the ground state energy from $H_\text{GNY}$. The rest of the Hamiltonian is
\bea
	V &=& \frac{1}{\sqrt{A}} \left[ \frac{1}{2} \pi^2 + \frac{\lambda \mathcal{A}^{\epsilon/2}}{4!}\varphi^4 + g_{\rm Y} \mathcal{A}^{\epsilon/4} \varphi \, \psi^{j \dagger} \gamma^0 \psi^{j} \right] \nn
	&& + \ \frac{\lambda \, \mathcal{A}^{\epsilon/2}}{8 \mathcal{A}} \varphi^2 \sum_{k \neq 0} \frac{\left[ a(-k) + a^{\dagger}(k) \right]\left[ a(k) + a^{\dagger}(-k) \right]}{|k|} \nn
	&& + \ \frac{g_{\rm Y} \, \mathcal{A}^{\epsilon/4}}{\sqrt{\mathcal{A}}} \varphi \sum_{k \neq 0} \sum_{s = 1}^{n_\text{D}/2} \left[ c^{\, j}_{s}(-k)b^{\, j}_s(k) + b^{j \, \dagger}_{s}(k) c^{\, j \dagger}_s(-k)  \right]  \nn
	&& + \ \cdots,
	\label{eq:intham}
\eea
where we only show the terms needed to obtain the leading loop corrections to the spectrum.

We now treat $V$ as a perturbation to $H_0$. The spectrum of $H_0$ is just the Fock spectrum, but crucially, every state in the unperturbed spectrum is infinitely degenerate. This is because the zero mode $\varphi$ does not appear, so we may multiply each eigenstate by an arbitrary normalizable function of $\varphi$ without changing the energy. Each state additionally has a $2^N$-fold degeneracy due to the fermionic zero modes, since we may arbitrarily choose $\psi^{j \dagger}_a \psi^j_a = 0,1$ for each value of $a$ and $j$.

We use an effective Hamiltonian method to treat $V$, which will describe the splitting of each Fock state due to interactions between the zero modes. We consider a degenerate subspace of $H_0$ with energy $\epsilon_0$, i.e., the set of states satisfying $H_0 | \alpha_0 \rangle = \epsilon_0 | \alpha_0 \rangle$. Then we construct an effective Hamiltonian which acts on this subspace, but whose eigenvalues are the exact eigenvalues, $H_\text{eff} | \alpha \rangle = E_{\alpha} | \alpha \rangle$, where $E_{\alpha} = \epsilon_0 + O(V)$ are the exact eigenvalues of $H$. This effective Hamiltonian may be obtained perturbatively in $V$, and at leading order it is given by~\cite{Bloch1958}
\beq
	H_\text{eff} = \epsilon_0 P_0 + P_0 V P_0 + P_0 V \frac{1 - P_0}{\epsilon_0 - H_0} V P_0 + \cdots,
	\label{eq:effham}
\eeq
where $P_0$ is the projection operator onto the degenerate subspace of interest. In this paper, we will only compute the effective Hamiltonian for the Fock vacuum, $|0\rangle$. In principle it is possible to obtain the effective Hamiltonian for any Fock state, but their structure becomes increasingly intricate at higher energies~\cite{Whitsitt2017}.

Taking $P_0 = |0\rangle \langle 0 |$, and combining Eqns~(\ref{eq:fockham}), (\ref{eq:intham}), and (\ref{eq:effham}), we obtain $H_\text{eff} = | 0 \rangle \langle 0 | h_{k=0}$, with
\bea
	h_{k=0} &=& \frac{1}{\sqrt{A}} \left[ \frac{1}{2} \pi^2 + \frac{\lambda \mathcal{A}^{\epsilon/2}}{4!}\varphi^4 + g_{\rm Y} \mathcal{A}^{\epsilon/4} \varphi \, \psi^{j \dagger} \gamma^0 \psi^{j} \right] \nn
	&& + \ \frac{\mathcal{A}^{\epsilon/2}}{8 \mathcal{A}} \varphi^2 \left( \lambda - 2 N g_{\rm Y}^2 \right) \sum_{k \neq 0} \frac{1}{|k|} + \cdots.
\eea
The sum in the second line of this expression is ultraviolet divergent. The evaluation of sums of this form using dimensional regularization is treated at length in Appendix C of Reference \onlinecite{Whitsitt2017}, so we simply quote the result:
\bea
	h_{k=0} &=& \frac{1}{\sqrt{A}} \left[ \frac{1}{2} \pi^2 + \frac{\lambda \mathcal{A}^{\epsilon/2}}{4!}\varphi^4 + g_{\rm Y} \mathcal{A}^{\epsilon/4} \varphi \, \psi^{j \dagger} \gamma^0 \psi^{j} \right] \nn
	&& + \ \frac{\mathcal{A}^{\epsilon/2}}{16 \pi \sqrt{\mathcal{A}}} \varphi^2 \left( \lambda - 2 N g_{\rm Y}^2 \right) \sqrt{\tau_2} f_{1/2}^{(3)}(\tau) + \cdots.
	\label{eq:vaceffham}
\eea
Here, we define the function
\bea
	f_{1/2}^{(3)}(\tau) &=& \int_1^{\infty} \frac{\dd \lambda}{\sqrt{\lambda}} \left[ \Theta(\lambda,\mathbf{\Omega}(\tau))^{3/2} - 1 \right] - \tau^{-3/2} - 2 \nn
	&& + \ \tau_2^{-3/2} \int_1^{\infty} \dd \lambda \left[ \Theta(\lambda,\mathbf{\Omega}(\tau)^{-1})^{3/2} - 1 \right],
	\label{eq:f12}
\eea
where the special function $\Theta(\lambda,\mathbf{\Omega})$, known as the two-dimensional Riemann Theta function, is defined as
\beq
	\Theta(\lambda,\mathbf{\Omega}) = \sum_{\mathbf{n} \in \mathbb{Z}^2} \exp\left( - \pi \lambda \, \mathbf{n}^{\mathrm{T}} \cdot \mathbf{\Omega} \cdot \mathbf{n} \right),
	\label{eq:rthet1}
\eeq
for a $2 \times 2$ matrix $\mathbf{\Omega}$. The matrices appearing in \eqref{eq:f12} are
\beq
	\mathbf{\Omega}(\tau) = \begin{pmatrix}
		| \tau|^2 & \tau_1 \\
		\tau_1 & 1
	\end{pmatrix}, \quad 
	\mathbf{\Omega}(\tau)^{-1} = \frac{1}{\tau_2^2}\begin{pmatrix}
		1 & -\tau_1 \\
		-\tau_1 & | \tau|^2
	\end{pmatrix}.
	\label{eq:rthet2}
\eeq

We now discuss the spectrum of $h_{k=0}$, which acts on the space of zero modes. The eigenfunctions are a product of a bosonic and fermionic part, $F[\varphi] \otimes | n_a^j \rangle$, where the zero-mode operators act as (temporarily using hats to distinguish operators from their eigenvalues)
\bea
	&\hat{\varphi} F[\varphi] = \varphi F[\varphi], \quad &\hat{\pi} F[\varphi] = -\I \frac{\partial}{\partial \varphi} F[\varphi], \nn
	&\psi^{j \dagger}_a\psi^{j}_a | n_a^{j} \rangle = n_a^{j}| n_a^{j} \rangle, \quad &n_a^{j} = 0, 1.
\eea
Focusing on the fermionic part of the Hilbert space, we can show that the effective Hamiltonian is symmetric under a full U($N$) symmetry group. This can be seen by choosing a basis such that $\gamma^0 = \mathrm{diag}(\mathbb{I},-\mathbb{I})$, after which the fermionic part of the Hamiltonian may be written 
\beq
	\psi^{j \dagger} \gamma^0 \psi^{j} = \sum_{a = 1}^{n_{\text{D}}/2} \left( \psi^{j \, \dagger}_{a} \psi^{j}_{a} - \psi^{j \, \dagger}_{a+n_\text{D}/2} \psi^{j}_{a+n_\text{D}/2}\right).
	\label{eq:enl_sym}
\eeq
Then by performing the transformation $\tilde{\psi}^{j}_{a+n_\text{D}/2} = \psi^{j \, \dagger}_{a+n_\text{D}/2}$, $\tilde{\psi}^{j \, \dagger}_{a+n_\text{D}/2} = \psi^{j}_{a+n_\text{D}/2}$, on the second term in the parenthesis, we have
\beq
	\psi^{j \dagger} \gamma^0 \psi^{j} \rightarrow \tilde{\psi}_a^{j \, \dagger} \tilde{\psi}^j_a - N/2,
	\label{eq:quad_psi}
\eeq
which is manifestly invariant under transformations of the ($\tilde{\psi}$) $\tilde{\psi}^{\dagger}$ fields as (anti-)fundamental vectors of U($N$). The emergent O($N$) symmetry of the chiral Ising CFTs noted at the end of Sec \ref{sec:qft} is the subgroup of this U($N$) obtained by taking purely real elements of the Lie group. The enlarged symmetry of the effective Hamiltonian compared to that in \eqref{eq:HGNY} occurs because the zero mode does not appear in the kinetic term, $\overline{\Psi}^j \cancel{\nabla}\Psi$, so the finite-momentum parts of $\Psi$ have less symmetry than the zero momentum part. Thus, we expect this extra symmetry of the zero-mode Hamiltonian to hold at all orders in perturbation theory. The results of Appendix \ref{app:largen} show that this emergent symmetry occurs in the $1/N$ and $\epsilon' = D - 2$ expansions as well.

Proceeding, we denote the above operator by $\hat{Q} \equiv \psi^{j \dagger} \gamma^0 \psi^{j}$, and its eigenvalues $Q$ take integer values in the range $Q \in [-N/2,N/2]$. The degeneracy of the eigenvalue $Q$ is
\beq
	\mathrm{deg}(Q) = \frac{N!}{(N/2 + Q)! (N/2 - Q)!}.
	\label{eq:q_deg}
\eeq
We now use \eqref{eq:gnyfixedpoint} to write the critical couplings as $g_{\mathrm{Y},\mathrm{ren}}^{\ast 2} = Y \epsilon$ and $\lambda_\text{ren}^{\ast} = U \epsilon$, where $Y$ and $U$ only depend on $N$. After the canonical transformation $\varphi \rightarrow \epsilon^{-1/6}$ and $\pi \rightarrow \epsilon^{1/6}$, our final form for the effective Hamiltonian is
\bea
	h^{(Q)}_{k=0} &=& \frac{\epsilon^{1/3}}{\sqrt{\mathcal{A}}} \left[ -\frac{1}{2} \frac{\partial^2}{\partial \varphi^2} + \frac{U}{4!}\varphi^4 + \sqrt{Y} Q \, \varphi \right] \nn
	&& + \ \frac{\epsilon^{2/3}}{16 \pi \sqrt{\mathcal{A}}} \, \varphi^2 \left( U - 2 N Y \right) \sqrt{\tau_2} f_{1/2}^{(3)}(\tau).
	\label{eq:effham_red}
\eea
This is the final version of the effective Hamiltonian that we will work with. The purpose of the canonical transformation was to make the $\epsilon$ dependence of the spectrum clear: the first line of \eqref{eq:effham_red} gives the leading $O(\epsilon^{1/3})$ contribution to the energy spectrum, and the second line (which required the computation of a one-loop diagram), gives the $O(\epsilon^{2/3})$ correction. The omitted terms in Eqns.~(\ref{eq:intham}) and (\ref{eq:effham}) can be shown to contribute only at higher orders in $\epsilon$ (we direct the interested reader to Ref.~\onlinecite{Whitsitt2017} for details on this point).

The lowest energies of the critical GNY torus spectrum are given by numerically solving the Hamiltonians in \eqref{eq:effham_red} for each $Q$, and for a given $Q$ the set of states obtained have a degeneracy given by \eqref{eq:q_deg}. Additionally, the spectrum of $h^{(Q)}_{k=0}$ is identical to the spectrum of $h^{(-Q)}_{k=0}$, where the bosonic part of the eigenfunctions are related by $F^{(Q)}[\varphi] = F^{(-Q)}[-\varphi]$. Thus, for a given $N$, we need to numerically solve the effective Hamiltonians for $Q = 0,1,...,N/2$. Since larger values of $|Q|$ lower the minimum of the potential, we expect that the ground states of the system to be given by the ground states of the $Q = \pm N/2$ sectors. From \eqref{eq:q_deg}, these two sectors are individually non-degenerate, and the resulting ground state is always exactly two-fold degenerate. We write the ground states as
\beq
	| \mathrm{GS}, \pm \rangle = F[\varphi] \, |N/2\rangle \pm F[-\varphi] \, |{-}N/2\rangle,
\eeq
where the $\pm$ index indicates the eigenvalue of this state under the $\mathbb{Z}_2$ symmetry $\varphi \rightarrow -\varphi$. The states $|\mathrm{GS},+ \rangle$ and $|\mathrm{GS},- \rangle$ correspond respectively to the levels denoted $1_T$ and $\sigma_T$ in earlier sections. The two ground states are \emph{exactly} degenerate in the scaling limit, although we expect the state $|\mathrm{GS},- \rangle$ to acquire a gap in any lattice realization of the transition due to non-universal corrections to scaling.

The first excited state then corresponds to the effective Hamiltonians with $Q = N/2 - 1$ and $Q = -N/2 + 1$. From \eqref{eq:q_deg}, this state has a total degeneracy of $2N$. The fermionic part of this state is obtained by acting on the ground state either with $\psi^{\dagger}_a$ for $1 \leq a \leq n_\text{D}/2$, or with $\psi_a$ for $n_\text{D}/2+1 \leq a \leq n_\text{D}$. These can be considered the particles ($n_f = 1$) or holes ($n_f = -1$) in the language of previous sections, and these states clearly correspond to those labelled $\psi_T$ earlier.

We note that the lowest-lying finite momentum states are obtained by constructing an effective Hamiltonian around the zeroth order finite momentum Fock states \cite{Schuler2016,Whitsitt2017}. In the present model, the energy of these states are not shifted from the zeroth-order value $|\mathbf{k}|$ ($v_F=1$ here) until order $\epsilon$, so at the order we are working they are unchanged compared to the Dirac CFT. This is in agreement with the small shift of this level seen in numerics, see Figure \ref{fig:spec_vs_kappa}.

We now detail the low-energy zero momentum spectrum for the $N=4$ case relevant to the model Hamiltonians $H_{\rm h}$ and $H_{\rm s}$, where we may relate each individual state with those obtained in numerics. A similar analysis of the spectrum may be done for any value of $N$.

\begin{table}
	\begin{tabular}{| c | c | c | c |}
		\hline 
		Level & $Q$ & deg. & $\sqrt{\mathcal{A}} \, E$ \\
		\hline
		$1_T$, $\sigma_T$ & $\pm 2$ & 2 & 0 \\
		\hline
		$\psi_T$ &  $\pm 1$ & 8 & $2.6033 \epsilon^{1/3} + 0.3287 \sqrt{\tau_2}f_{1/2}^{(3)}(\tau) \epsilon^{2/3}$  \\
		\hline
		C, K, $(2\psi_T)$ &  $0$ & 6 & $3.7443 \epsilon^{1/3} + 0.5698 \sqrt{\tau_2}f_{1/2}^{(3)}(\tau) \epsilon^{2/3}$  \\
		\hline
		$\epsilon_T$, $\sigma_T'$ &  $\pm 2$ & 2 & $4.4713 \epsilon^{1/3} + 0.1183 \sqrt{\tau_2} f_{1/2}^{(3)}(\tau) \epsilon^{2/3}$  \\ 
		\hline
	\end{tabular}
    \caption{Low-lying spectrum of the $N=4$ GNY model on the torus, from numerically computing the strong-coupling expansion in the text. We measure energies with respect to the ground state. The states are labeled by the eigenvalue $Q$ defined in the text, and their degeneracy (deg.) is given. The labelling of the levels is used to agree with that used in previous sections, where C and K refer to Chern and Kekule respectively.}
	\label{tab:eps_spectrum}
\end{table}

\subsection{The case of $N = 4$}

Specializing to the case with four degrees of freedom, we need to solve for the lowest eigenvalues of \eqref{eq:effham_red} for $Q = 0, 1, 2$. We obtain the spectrum by first computing the eigenfunctions and eigenvalues of the $O(\epsilon^{1/3})$ part of the spectrum numerically, giving us the leading-order contribution to the energy and eigenfunctions. We then compute the $O(\epsilon^{2/3})$ contribution from these eigenfunctions using ordinary first-order perturbation theory. The results of this computation are shown in Tab.~\ref{tab:eps_spectrum}. In Tab.~\ref{tab:eps_numspectrum} we list the explicit numbers for the torus shapes considered in this manuscript.

By looking at the transformation of the eigenfunctions under the symmetries in Section \ref{sec:symmetry}, we can explicitly relate these states to the $\kappa=0$ states enumerated in the numerical simulations of previous sections. In \figref{fig:spec_vs_kappa}(b), we compare the resulting spectrum from $\epsilon$-expansion to the one obtained from numerics. We observe a favorable agreement between the two methods. In particular, the sequence of the eigenstates' quantum numbers and degeneracies (quasi-degeneracies in numerics, see below) are identical. Also, the relative energy gaps in the $\kappa=0$ sector are similar, i.e. we observe large gaps between the $Q=\pm2$ levels to the other states, while the $Q=0$ and $Q=\pm1$ levels are very close in energy. Quantitatively, the $\epsilon$-expansion underestimates the gaps observed in numerics, which we relate to be mainly an artefact of the low-order expansion.
 
The degeneracies of the torus spectrum levels because of the emergent O($N$) symmetry of the chiral Ising CFT [see Secs.~\ref{sec:eps_expansion_structure},~\ref{sec:qft}] is a highly nontrivial prediction of the field theory, suggesting that the further splitting seen between these levels in numerics is non-universal and an artefact of corrections to scaling in explicit lattice realizations.

We also note that the energy depends on the shape of the torus rather weakly [Tab.~\ref{tab:eps_numspectrum}], but that the levels for the square torus are slightly higher than those for the triangular torus.

\begin{table}
	\begin{tabular}{| c | c | c | c | c |}
		\hline 
		Level & $Q$ & deg. & $\frac{3}{4 \pi}\xi^{\text{CI}}_{{\rm h}}$ & $\frac{3}{4 \pi}\xi^{\text{CI}}_{{\rm s}}$ \\
		\hline
		$1_T$, $\sigma_T$ & $\pm 2$ & 2 & 0 & 0 \\
		\hline
		  $\psi_T$ &  $\pm 1$ & 8 & 0.246 & 0.247 \\
		\hline
		  C, K, $(2\psi_T)$ &  $0$ & 6 & 0.312 & 0.315 \\
		\hline
		   $\epsilon_T$, $\sigma_T'$ & $\pm 2$ & 2 & 0.612 & 0.613 \\ 
		\hline
	\end{tabular}
    \caption{Low-lying spectrum of the $N=4$ GNY model for particular torus shapes considered in numerics, written in terms of the $\xi^{\text{CI}}$ defined in Section \ref{sec:gny_cft}. We measure energies with respect to the ground state, extrapolated to $\epsilon=1$. Here we give the energy spectrum for modular parameters $\tau = \exp(i \pi/3)$ and $\tau = i$ appropriate to the honeycomb and square lattices respectively. }
	\label{tab:eps_numspectrum}
\end{table}

As discussed earlier, the chiral Ising CFT with $N$ degrees of freedom appears to always flow to a fixed point with full O($N$) symmetry in perturbation theory, even when the original field theory does not possess this symmetry. We have already noted how this symmetry appears in the torus spectrum below \eqref{eq:enl_sym}, where it is a subgroup of a larger SU($N$) symmetry. Therefore, we may classify the states in Table \ref{tab:eps_spectrum} by their representations under these symmetry groups. From this perspective, the large degeneracies of the torus spectrum may be related to the large emergent symmetry of the CFT. Obtaining the relevant irreducible representations for a given state is easiest when $\hat{Q}$ is written in the form of \eqref{eq:quad_psi}, where the states are given by acting on the lowest-$Q$ state by antisymmetrized products of the SU($N$) vectors $\tilde{\psi}^{j \, \dagger}_a$. In this way, we see that the $Q = \pm 2$ states are SU(4) singlets, the eight $Q = \pm 1$ states are two inequivalent SU(4) vectors, and the six-fold degenerate $Q=0$ states transform into each other as an antisymmetric SU(4) tensor. The enumerations of multiplets and their degeneracies is not altered if we instead consider the O(4) subgroup of SU(4).

\section{Conclusions \& Outlook}
\label{sec:conclusion}

In this manuscript we have shown how to calculate the critical torus energy spectrum for the chiral Ising universality class of the GNY theory with $n_{\mathrm{D}}=4$ spinor components in $D=(2+1)$ dimensions from the investigation of strongly interacting fermionic tight-binding models. We have computed the low-energy spectrum on finite-size clusters on different spatial torus geometries (honeycomb vs. square models) using exact diagonalization and quantum Monte Carlo approaches, which complement each other particularly well for this task. We have extrapolated the finite-size results to the thermodynamic limit to obtain the critical torus energy spectrum which serves as a unique fingerprint of the QCP's universality class. 

Furthermore, we have calculated the critical torus energy spectrum for the chiral Ising universality class using the perturbative expansion in $\epsilon = 4 - D$. This analytical approach shows a good qualitative agreement with our numerical results. In particular it predicts non-trivial degeneracies of levels which we also observe in numerics after extrapolation to the thermodynamic limit. This validates the description of the QCPs in the lattice models as chiral Ising critical points of the GNY theory. The $\epsilon$-expansion results also suggest, that the critical torus spectrum of GNY theories only depends on the total number of fermionic degrees of freedom $N=n_\text{D} N_f$, instead of depending on $n_\text{D}$ and $N_f$ individually.

We also observe that the finite-size clusters used to approach the thermodynamic limit split into families with distinct critical torus spectra. These families can be distinguished by the properties of the clusters momentum space; one family of clusters has the Dirac points in their momentum space and, in the field theory, correspond to periodic boundary conditions of the fermionic and bosonic fields. The other family does not have the Dirac points and describes twisted boundary conditions for the fermionic fields, while the bosonic field remains unaltered. The critical torus spectrum is very different for the different families.

Furthermore, we have computed the renormalization of the Fermi velocity due to the interactions between the fermions in the SM phase, which we derive from the renormalization of the torus energy level of a single fermion mode. We have shown that the so-obtained approximately linear velocity renormalization also describes the behaviour of other energy levels in the SM phase. Assuming that the Fermi velocity behaves continuously also at the critical point, we extrapolate the linear renormalization to the critical point, and obtain an approximately 35\% increase at criticality compared to the non-interacting case. Additionally, we have investigated the crossover behaviour between the chiral Ising and the Dirac critical points for finite-size systems. We point out, that this crossover behaviour can lead to bad estimators for the Fermi velocity renormalization.

We hope that this work further strengthens the interpretation of the critical torus energy spectrum as a universal fingerprint of quantum critical points, that is, as we have seen here, capable of detecting the coupling of bosonic fields to fermionic spinors. We anticipate that this work inspires future research on the critical torus spectrum for chiral Ising models using different methods, a different number of spinor components $N$, or on chiral XY and chiral Heisenberg models where the spinors are coupled to a continuous O(2)/O(3)-order parameter. Recently, it was shown how to create fermionic tight-binding models with a single Dirac cone that is exactly linear in the entire Brillouin zone of the finite-size system~\cite{Lang2019}. Such systems could also be very beneficial to study chiral universality classes, because the portion of the Brillouin zone showing Dirac physics is much larger than in the models considered within this chapter, and finite-size extrapolation might become easier. Finally we believe our results complete an important step towards a {\em quantitative} understanding of the torus energy spectrum of QED$_3$-like theories, believed to describe quantum spin liquids on the Kagome and the triangular lattice~\cite{Ran2007,Iqbal2013,He2017,Song2019}.

\begin{acknowledgments}
MS, TCL, and AML acknowledge support by the Austrian Science Fund for project SFB FoQus (F-4018) and DFG-FOR1807 (I-2868). MS acknowledges support by the Austrian Science Fund (FWF) through Grant No. P 31701-N27. SH and StW acknowledge support by the Deutsche Forschungsgemeinschaft (DFG) under project number RGT 1995. SeW acknowledges support from the NIST NRC Postdoctoral Associateship award. The computational results presented have been achieved in part using the Vienna Scientific Cluster (VSC). This work was supported by the Austrian Ministry of Science BMWF as part of the UniInfrastrukturprogramm of the Focal Point Scientific Computing at the University of Innsbruck. We thank the IT Center at RWTH Aachen University and the JSC J\"ulich for access to computing time through JARA-HPC.
\end{acknowledgments}

\section*{Appendix}
\appendix

\section{Operators for Gap Estimation with QMC simulations}
\label{app:qmc_operators}

\begin{table}[ht]
\begin{tabular}{|c|c|c|c|c|c|c|c|c|}
\hline
Level & PH & $n_f$  & $\mathbf{k}$ & $C_6$ & $C_3$ & $C_2$ & $I_x$ & $I_y$ \\
\hline
$1_T,\epsilon_T$ & $1$& $0$  & ${0}$ & $1$ & $1$ & $1$ & $1$ & $1$ \\
\hline
$\sigma_T, \sigma_T^\prime$ & $-1$ & $0$  & ${0}$ & $1$ & $-1$& $-1$ & $-1$ & $1$ \\
\hline
Kekule & $1$ & $0$  & $\mathbf{K}$ & $1$ & -- & -- & $1$ & --  \\
\hline
Chern & $-1$ & $0$  & ${0}$ & -- & -- & $-1$ & $-1$ & $1$ \\
\hline
$\psi_T$ & -- & $1$  & $\mathbf{K}$ & -- & -- & -- & -- & -- \\
\hline
$(2\psi)_T$ & -- & $2$  & ${0}$ & -- & -- & -- & -- & -- \\
\hline
\end{tabular}
\caption{Symmetry properties of the operators connecting the ground state to the relevant excited states for the honeycomb lattice. The tables denotes the action under particle-hole transformation PH, the fermion sector $n_f$ and the momentum $\mathbf{k}$, as well as the action under rotations about $60^{\circ}$ ($C_6$) and $120^{\circ}$ ($C_3$), lattice inversion ($C_2$), and vertical ($I_x$) and horizontal ($I_y$) mirror reflection. $1\ (-1)$ indicates a vanishing commutator (anti-commutator), and -- corresponds to an otherwise broken symmetry. In the case of degenerate excited states operators with different symmetries can be allowed, which connect to superimposed states of the degenerate subspace.}\label{tab:sym}
\end{table}

The LCT-INT algorithm~\cite{Wang2015} used in this work  makes explicit use of a weak coupling expansion of the partition function in terms of the interacting part of the Hamiltonian. As a result imaginary time displaced expectations values of the form in \eqref{eq:OOdag} can be expressed as expectation values of the free Hamiltonian. One can therefore employ the Wick theorem to calculate a Monte Carlo estimator of \eqref{eq:OOdag} using Green functions of the form
\begin{align}
	G_{ij}(\tau) = \left \langle c_i (\tau)\, c^\dag_j (0) \right \rangle \label{eq:G_tau}\,,
\end{align}
which are calculated during the LCT-INT sampling process.

\subsection{The $\sigma_T$ Level}

The staggered density operator
\begin{align}
	\hat{O}_{\sigma_T} = \sum_{i} \left( -1 \right)^i \left( n_i - \frac{1}{2} \right) \label{eq:O_cdw}
\end{align}
corresponds to the order parameter of the commensurate charge-density-wave of the bipartite lattice [see \figref{fig:bonds}(b)]. The operator is anti-symmetric under particle-hole transformation and connects the $\mathbb{Z}_2$-even and -odd quasi-degenerate lowest energy states with each other. Furthermore, this operator provides an overlap of the ground state with the energetically higher $\sigma_T^\prime$ level.

\subsection{The $\psi_T$ and $(\Psi + d)_T$ Levels}

The $n_f = 0$ sector can be connected to the $n_f = 1$ ($n_f = -1$) sector by operators that create (annihilate) a fermion with a certain momentum. For lattice clusters that contain the Dirac point, the lowest excited state of the $n_f = 1$ sector is connected to the $n_f = 0$ ground state by the operator
\begin{align}
	\hat{O}_{\psi_T} = c_{\mathbf{K}}^{\dag} = \sum_{i} \E^{\I \mathbf{K} \cdot \mathbf{r}_i} c_i^{\dag},
\end{align}
which creates a fermion at the Dirac point $\mathbf{K}$. Note that the eight-fold degeneracy of the $\sigma_T$ level follows from the valley, orbital and particle-hole degeneracy, $2_{\text{valley}} \times 2_{\text{orbital}} \times 2_{\text{PH}} = 8$.

Overlap to higher excited states can then be achieved by creating a fermion with the $n$-th closest momentum to the Dirac point,
\begin{align}
	\hat{O}_{(\psi + n d)_T} = c_{\mathbf{K}+\mathbf{q}}^{\dag} = \sum_{i} \E^{\I (\mathbf{K}+\mathbf{q}) \cdot \mathbf{r}_i} c_i^{\dag}\,,
\end{align}
where the momentum $\mathbf{q}$ is chosen accordingly on the lattice cluster.

\subsection{The $2 \psi_T$ Level}

The $(2\psi)_T$ level corresponds to the lowest excited state in the $n_f = \pm 2$ sectors. The ground state can be connected to them by operators that create two fermions, one with momentum $\mathbf{K}$ and one with $-\mathbf{K}$,
\begin{align}
	\hat{O}_{2 \psi_T} = c_{\mathbf{K}}^{\dag} c_{\mathbf{-K}}^{\dag} = \sum_{ij} \E^{\I \mathbf{K} \cdot (\mathbf{r}_i - \mathbf{r}_j)} c_i^{\dag} c_j^{\dag}\,.
\end{align}
Note that the total momentum is zero. In this case, the valley and orbital degeneracies become redundant, and the two-fold degeneracy of the $2 \psi_T$ level follows directly from particle-hole symmetry.

\subsection{The $\epsilon_T$ Level}

The $\epsilon_T$ level corresponds to the first excited state with identical symmetries as the ground state. In order to connect states within the same symmetry sector, possible operators must commute with all symmetry operations. A suitable choice is therefore given by either part of the Hamiltonian,
\begin{align}
	\hat{O}_{\epsilon_T}^{(t)} &= -t \sum_{\langle ij \rangle} \left( c_i^{\dag} c^{\pdag}_j + c_j^{\dag} c^{\pdag}_i \right), \\
	\hat{O}_{\epsilon_T}^{(V)} &= V \sum_{\langle ij \rangle} \left( \hat{n}_i - \frac{1}{2} \right) \left( \hat{n}_j - \frac{1}{2} \right).
\end{align}
Since both operators have a finite ground state expectation value, one has to extract the gap $\Delta_{\epsilon_T}$ using the formula
\begin{align}
	\left \langle \hat{O}_{\epsilon_T}(\tau) \hat{O}_{\epsilon_T}^\dag \right \rangle \sim \left| \left\langle \Psi_0 \right| \hat{O}_{\epsilon_T} \left| \Psi_{\epsilon_T} \right\rangle\right|^2 \E^{- \Delta_{\epsilon_T} \tau} + \left| \left\langle \Psi_0 \right| \hat{O}_{\epsilon_T} \left| \Psi_0 \right\rangle\right|^2\,.
\end{align}
Note that the operator $\hat{O}_{\epsilon_T}^{(V)}$ is related to the weak coupling expansion used in LCT-INT. One can therefore calculate its correlation function, as well as the fidelity susceptibility, from the distribution of interaction vertices during the Monte Carlo sampling~\cite{Wang2015}.

\subsection{The Kekule Level}

The Kekule level corresponds to the lowest excited states in the $n_f = 0$ sector with momentum $\mathbf{K}$ and identical particle-hole parity as the ground state. This level is two-fold degenerate due to the valley degeneracy. Possible operators can be constructed from the Kekule bond pattern, which itself is 3-fold degenerate on the honeycomb lattice ($K_1$, $K_2$ and $K_3$) [see \figref{fig:bonds}(c)]. The Kekule pattern features an enlarged unit cell, which in reciprocal space corresponds to the momentum at the Dirac point. Because of the reduced lattice symmetry at finite momenta, states with momentum $\mathbf{K}$ do not have a well defined inversion parity. Nevertheless, one can choose to construct the Kekule operators such that they are (anti-)symmetric under lattice inversion. In this case the operators do not have well defined momenta, and provide overlap of the ground state with states of momentum $\mathbf{K}$ as well as $-\mathbf{K}$,
\begin{align}
	\hat{O}_{\text{Kekule}}^{(+)} = \sum_{\langle ij \rangle \in K_1} &\left( c_i^{\dag} c^{\pdag}_j + c_j^{\dag} c^{\pdag}_i \right) + \sum_{\langle ij \rangle \in K_2} \left( c_i^{\dag} c^{\pdag}_j + c_j^{\dag} c^{\pdag}_i \right) \nonumber \\
	 - 2 \sum_{\langle ij \rangle \in K_3} &\left( c_i^{\dag} c^{\pdag}_j + c_j^{\dag} c^{\pdag}_i \right), \\
\hat{O}_{\text{Kekule}}^{(-)} = \sum_{\langle ij \rangle \in K_1} &\left( c_i^{\dag} c^{\pdag}_j + c_j^{\dag} c^{\pdag}_i \right) + \sum_{\langle ij \rangle \in K_2} \left( c_i^{\dag} c^{\pdag}_j + c_j^{\dag} c^{\pdag}_i \right).
\end{align}
Note that both $\hat{O}_{\text{Kekule}}^{(+)}$ and $\hat{O}_{\text{Kekule}}^{(-)}$ transform symmetric under the $\mathbb{Z}_2$ particle-hole transformation.

\subsection{The Chern Level}

\begin{figure}[htb]
	\centering
	\includegraphics[height=6cm]{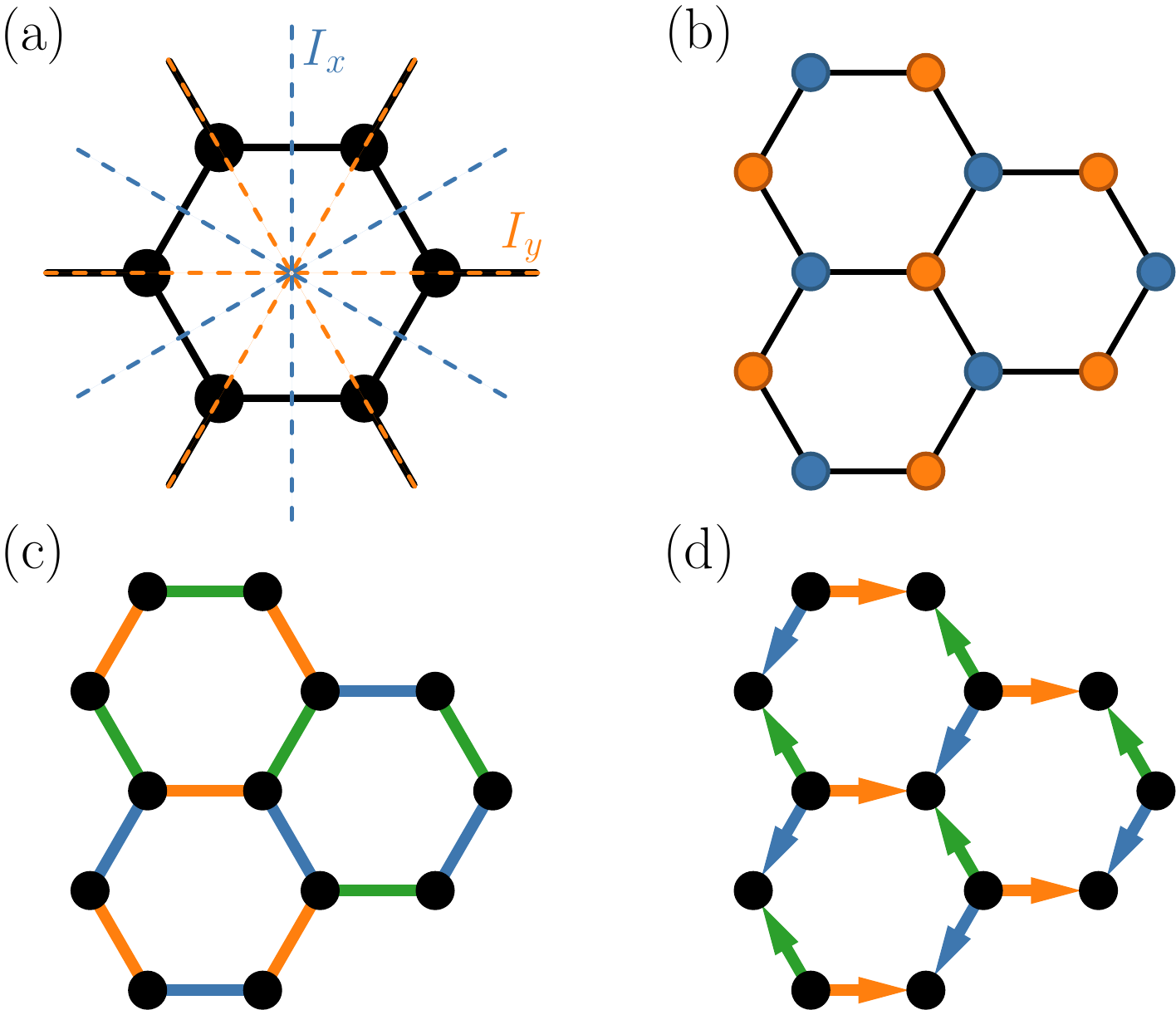}
	\caption{Real space illustration of the point group symmetry of the honeycomb lattice (a), as well as the operator representations used for the $\sigma_T$ (b), Kekule (c) and Chern level (d). In panel (a)  the blue (orange) dotted lines denote the axes of vertical (horizontal) mirror reflection, while the fixed point of lattice rotations is the center of the hexagon. Panel (b) shows the anti-symmetric sub-lattice modulation of \eqref{eq:O_cdw} with the standard two-site unit cell of the honeycomb lattice. Panel (c) depicts the three distinct Kekule patterns $K_1$, $K_2$ and $K_3$ of the honeycomb lattice, and panel (d) depicts the current pattern of the bonds $N_1$, $N_2$ and $N_3$ used in \eqref{eq:O_chern}.}
	\label{fig:bonds}
\end{figure}

The Chern level denotes the lowest excited states in the $n_f = 0$ sector with zero momentum and opposite particle-hole parity as the ground state. These states are two-fold degenerate and transform according to a two-dimensional irreducible representation of the point group. They can be connected to the ground state by current operators that are anti-symmetric under particle-hole transformation and break rotational symmetry [see \figref{fig:bonds}(d)], such as
\begin{align}
\hat{O}_{\text{Chern}} = \I \sum_{\langle ij \rangle \in N_1} &\left( c_i^{\dag} c^{\pdag}_j - c_j^{\dag} c^{\pdag}_i \right) + \frac{\I}{2} \sum_{\langle ij \rangle \in N_2} \left( c_i^{\dag} c^{\pdag}_j - c_j^{\dag} c^{\pdag}_i \right) \nonumber \\
+ \frac{\I}{2} \sum_{\langle ij \rangle \in N_3} &\left( c_i^{\dag} c^{\pdag}_j - c_j^{\dag} c^{\pdag}_i \right) \label{eq:O_chern}.
\end{align}

\section{Critical Torus Spectrum for $H_{\rm s}$ }
\label{app:sq_numerics}

\begin{figure*}[htb]
	\centering
	\includegraphics[width=\textwidth]{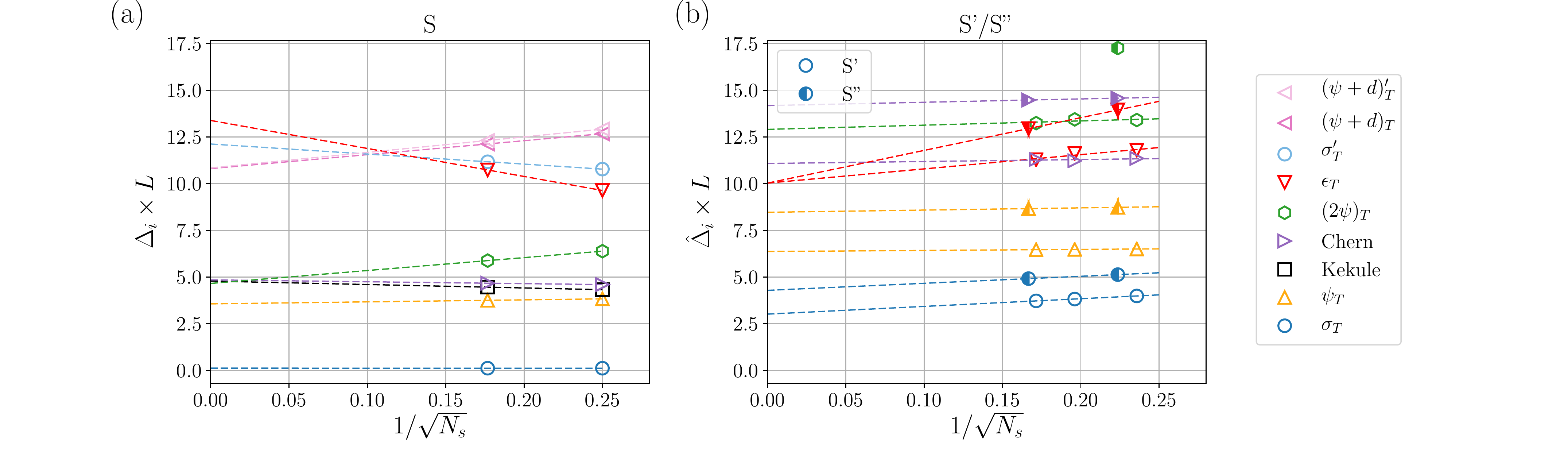}
	\caption{Extrapolation of the critical torus spectrum on the $\pi$-flux square lattice for clusters that contain the Dirac points (a), or not (b). In (b) there are two classes of lattices, denoted S' (empty symbols) and S'' (half-filled symbols). The label for the states (see legend on the left) is indicated by the color and shape of the symbols in the right panel. See \figref{fig:hc_extrap} for a comparison.}
	\label{fig:sq_extrap}
\end{figure*}

In this appendix we analyze the chiral Ising QCP of the model \eqref{eq:sqmodel} on the $\pi$-flux square lattice. For this case, we accessed finite-size data only from ED and we show the finite-size extrapolations of the most important low-energy levels in \figref{fig:sq_extrap}. For the ED calculations for $H_{\rm s}$, we used the uniform gauge choice $\theta_{ij} = \pi/4$, $\theta_{ji} = -\theta_{ij}$, in order to assure a four-fold rotational symmetry of the Hamiltonian.  The extrapolations of model \eqref{eq:hcmodel} on the honeycomb lattice [see~\figref{fig:hc_extrap}] show that the extrapolations of ED data alone typically give rather good agreement with the extrapolations of much larger clusters from QMC data. We, thus, assume that the extrapolations of the square-lattice model also give satisfactory qualitative estimates for the critical torus spectrum on the $\pi$-flux square lattice. 

The critical torus spectrum for the clusters that contain the Dirac points [see left panel in \figref{fig:sq_extrap}] shows a very similar structure to the one of the honeycomb lattice, as it is also illustrated in \figref{fig:overview} in the main text. Again, the $\sigma_T$ level shows a very low energy gap, and the Kekule, Chern and $(2\psi)_T$ levels are very close to each other, forming a $6$-fold (nearly) degenerate level in the thermodynamic limit. Furthermore, both the $\epsilon_T$ and $\sigma_T^\prime$ levels, as well as the $(\psi+d)_T$ and $(\psi+d)_T^{\prime}$ levels, become nearly degenerate, as was observed on the honeycomb lattice, and suggested from the $\epsilon$-expansion.

In contrast to the honeycomb case, the sequence of square lattices without Dirac points (and full four-fold rotational $C_4$ symmetry) separates into two families, which we denote as S' and S'', respectively [see right panel in \figref{fig:sq_extrap}]. These two families are distinguished by the distance $||\mathbf{X} - \mathbf{k}_\text{min}|| L$ of the closest momentum point $\mathbf{k}_\text{min}$ to the Dirac point $\mathbf{X}$ [see \figref{fig:lattice_families}]. The scaled energy gap $\Delta_i \times L$ of, for example, the single particle mode $n_f=\pm1$ is, thus, different for both families in the non-interacting case, $V=0$, which develops into a distinct critical spectrum for clusters of family S' and S''. While the finite-size extrapolated $\epsilon_T$ levels are very similar for both of those families, the other levels are rescaled, but the sequence remains almost unchanged and is comparable with the results from the honeycomb lattice.

\begin{figure*}[htb]
	\centering
	\includegraphics[width=\textwidth]{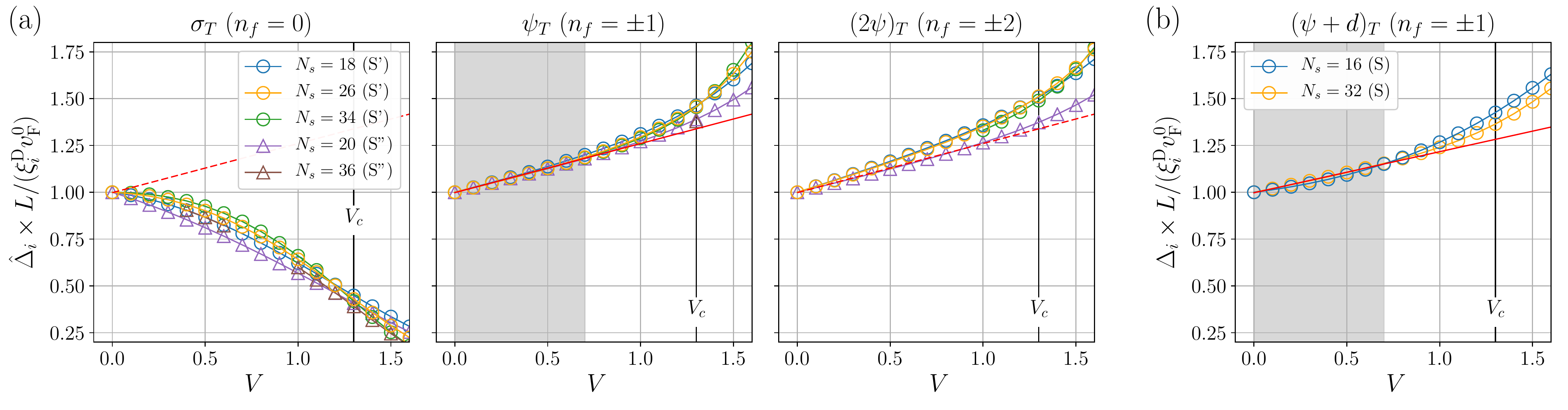}
	\caption{Renormalization of the Fermi velocity for the $\pi$-flux square lattice model. Here we use the not-extrapolated ${N_s=34}$ site data for the extrapolations in (a) [${N_s=32}$ in (b)]. The shaded region indicates the fitting window. See \figref{fig:hc_vf_ren} for a comparison.}
	\label{fig:sq_vf_ren}
\end{figure*}

We can also attempt to compute the renormalization of the Fermi velocity with the interaction strength $V$ in the SM phase $V<V_c$ for the $\pi$-flux square lattice model in an analogous way to the honeycomb case [see Section~\ref{sec:hc_numerics_vfren}]. We again choose the single-fermion level at $\mathbf{k}_\text{min}$ to derive the velocity renormalization, because it shows the smallest finite-size effects. We here use the largest finite size cluster to estimate the Fermi-velocity renormalization, since an extrapolation of the energy gaps to the thermodynamic limit for each $V<V_c$ could not be reliably obtained for this model. Like in the case of the honeycomb lattice, the renormalization can be well approximated by a linear function, and the behaviour of other characteristic levels, like the two-fermion level $n_f=\pm2$, is consistent [see right panel in \figref{fig:sq_vf_ren}(a)]. The lowest $n_f=0$ excitation ($\sigma_T$) shows strong finite-size effects since it is strongly influenced by the crossover effects described in Sec.~\ref{sec:cross}, and, without QMC data, the available system-sizes are too small to read off the Fermi velocity renormalization from this level [see left panel in \figref{fig:sq_vf_ren}(a)]. The family of clusters that contain the Dirac points also yields a compatible renormalization of the Fermi velocity [see \figref{fig:sq_vf_ren}(b)]. At criticality, the Fermi velocity is increased by approximately $30\%$ compared to the non-interacting value $v_\text{F}^0$.

\section{Derivation of Quantum Field Theories from Model Hamiltonians}
\label{app:micromap}

In this Appendix, we explicitly derive the GNY and GN field theories directly from the microscopic model Hamiltonians of Section \ref{sec:model_ham}. We will mostly focus on the honeycomb lattice Hamiltonian $H_{\rm h}$; the corresponding derivation for $H_{\rm s}$ follows identical steps, and the only differences are unimportant numerical factors. This explicit computation is done to aid mapping the microscopic symmetries of our model Hamiltonians to those in the GNY and GN models, outlined in Section \ref{sec:symmetry}.

We first recall the exact solution of the $V=0$ limit of the honeycomb model $H_{\rm h}$~\cite{Semenoff1984} in order to fix notation before adding interactions. The honeycomb model has two sites per unit cell, and one may write the quadratic part of the Hamiltonian \eqref{eq:hcmodel} as
\bea
	H_{{\rm h},0} = - \sum_{\mathbf{r}_i} \sum_{j = 1}^3 \left( c^{\dagger}_{\mathbf{r}_i,A} c^{\phantom{\dagger}}_{\mathbf{r}_i + \mathbf{s}_j,B} + \mathrm{h.c.} \right).
\eea
Here, the sum $\sum_{\mathbf{r}_i}$ is over all unit cells in the lattice, and ${\mathbf{s}_1 = (1,0)}$, ${\mathbf{s}_2 = (-1/2,\sqrt{3}/2)}$, and ${\mathbf{s}_3 = (-1/2,-\sqrt{3}/2)}$ are the unit vectors connecting nearest-neighbors on the honeycomb lattice (we take the lattice spacing and the coupling $t$ to unity). We define the Fourier transform on the infinite honeycomb lattice by
\beq
	\begin{pmatrix}
		c_{\mathbf{r}_i,A} \\
		c_{\mathbf{r}_i,B}
	\end{pmatrix} = \int_{\mathrm{BZ}} \frac{\dd^2 k}{\mathcal{A}_{\mathrm{BZ}}} \E^{\I \mathbf{k} \cdot \mathbf{r}_i} \begin{pmatrix}
		c_{\mathbf{k},A}\, \E^{- i \mathbf{k} \cdot \mathbf{s}_1/2} \\
		c_{\mathbf{k},B} \,\E^{ i \mathbf{k} \cdot \mathbf{s}_1/2}
	\end{pmatrix},
	\label{eq:FTdef}
\eeq
where the integral is over the Brillouin zone, and $\mathcal{A}_{\mathrm{BZ}} = 8 \pi^2/3\sqrt{3}$ is its area. This transforms the Hamiltonian to
\beq
	H_{{\rm h},0} = \int_{\mathrm{BZ}} \frac{\dd^2 k}{\mathcal{A}_{\mathrm{BZ}}} \left( h(\mathbf{k}) c^{\dagger}_{\mathbf{k},A} c^{\phantom{\dagger}}_{\mathbf{k},B} + \mathrm{h.c.}  \right),
	\label{eq:honeyc_exact}
\eeq
with $h(\mathbf{k}) = -\sum_{i=1}^3 \exp(\I \mathbf{k} \cdot (\mathbf{s}_i + \mathbf{s}_1))$. The resulting tight-binding spectrum is $E(\mathbf{k}) = \pm |h(\mathbf{k})|$, and it is pictured in Fig.~\ref{fig:gnymodels}. It vanishes at two inequivalent points in the Brillouin zone, $\mathbf{K}$ and $\mathbf{K}'$. Defining an appropriate four-spinor,
\beq
	\begin{pmatrix}
		\Psi_1(\mathbf{k}) \\ \Psi_2(\mathbf{k}) \\  \Psi_3(\mathbf{k}) \\\Psi_4(\mathbf{k}) \\ 
	\end{pmatrix} = 
	\begin{pmatrix}
		\kappa \, c_{\mathbf{K} + \mathbf{k},A} \\ \kappa^{\ast} c_{\mathbf{K} + \mathbf{k},B} \\  \kappa \, c_{\mathbf{K}' + \mathbf{k},A} \\ \kappa^{\ast} c_{\mathbf{K}' + \mathbf{k},B} \\ 
	\end{pmatrix},
	\label{eq:spinordef}
\eeq 
with $\kappa = (2 \pi)\E^{-i \pi/6}/\sqrt{\mathcal{A}_{\mathrm{BZ}}}$, the low energy effective Hamiltonian for $V=0$ becomes
\bea
	H_{{\rm h},0} &=& v_\text{F}\int \frac{\dd^2 k}{(2 \pi)^2} \overline{\Psi}(\mathbf{k}) \,\I \left[ \gamma^1 k_x + \gamma^2 k_y \right] \Psi(\mathbf{k}) \nn
	&=& v_\text{F} \int \dd^2 x \, \overline{\Psi}(\mathbf{x}) \cancel{\nabla} \Psi(\mathbf{x})\,,
\eea
where $v_\text{F} = 3/2$, and the explicit representation of the gamma matrices being used is given in \eqref{eq:gamma_rep}.

We now consider interactions, $V>0$. We first define the scalar order parameter:
\beq
	\Phi(\mathbf{r}_i) \equiv n_{\mathbf{r}_i,A} - n_{\mathbf{r}_i,B}.
\eeq
This measures the difference in density on the two sublattices, so its expectation value should vanish in the SM phase, while it saturates to $\pm1$ deep in the ordered phase. We note the following identity,
\bea
	\sum_{\mathbf{r}_i} \Phi(\mathbf{r}_i)^2 &=& - 2 \int_{\mathrm{BZ}} \frac{\dd^2 k_1 \, \dd^2 k_2 \, \dd^2 q}{\left(\mathcal{A}_{\mathrm{BZ}}\right)^3} c^{\dagger}_{\mathbf{k}_1 + \mathbf{q},A} c^{\dagger}_{\mathbf{k}_2 - \mathbf{q},B} c^{\phantom{\dagger}}_{\mathbf{k}_2,B} c^{\phantom{\dagger}}_{\mathbf{k}_1,A}\nn
&& + \  \int_{\mathrm{BZ}} \frac{\dd^2 k}{\mathcal{A}_{\mathrm{BZ}}} \left( c^{\dagger}_{\mathbf{k},A}c^{\phantom{\dagger}}_{\mathbf{k},A} + c^{\dagger}_{\mathbf{k},B}c^{\phantom{\dagger}}_{\mathbf{k},B} \right)\,.
	\label{eq:orderparameter}
\eea
We compare this to the interaction term in $H_{\rm h}$:
\bea
	H_{{\rm h},{\rm int}} &=& V \sum_{\nnb} \left(n_i - \frac{1}{2} \right)
    \left(n_j - \frac{1}{2} \right) \nn
    &=& V \int_{\mathrm{BZ}} \frac{\dd^2 k_1 \, \dd^2 k_2 \, \dd^2 q}{\left(\mathcal{A}_{\mathrm{BZ}}\right)^3} h(\mathbf{q}) c^{\dagger}_{\mathbf{k}_1 + \mathbf{q},A} c^{\dagger}_{\mathbf{k}_2 - \mathbf{q},B} c^{\phantom{\dagger}}_{\mathbf{k}_2,B} c^{\phantom{\dagger}}_{\mathbf{k}_1,A} \nn
	&& - \frac{3V}{2} \int_{\mathrm{BZ}} \frac{\dd^2 k}{\mathcal{A}_{\mathrm{BZ}}} \left( c^{\dagger}_{\mathbf{k},A}c^{\phantom{\dagger}}_{\mathbf{k},A} + c^{\dagger}_{\mathbf{k},B}c^{\phantom{\dagger}}_{\mathbf{k},B} \right),
	\label{eq:hcint_k}
\eea
where $h(\mathbf{q})$ is the same function appearing in \eqref{eq:honeyc_exact}, and we have dropped a constant. In the low-energy limit, we are only interested in the regions of these integrals where the fermions may be expanded in terms of the $\Psi$ fields of \eqref{eq:spinordef}. Analyzing the first line of \eqref{eq:hcint_k}, we find that the only nonzero values of $\mathbf{q}$ which allow this expansion are $\mathbf{q} = \mathbf{K}$, $\mathbf{K}'$, but these are precisely where $h(\mathbf{q}) = 0$, so these regions do not contribute. Thus, it suffices to expand the integrand near $\mathbf{q} = 0$, where  $h(\mathbf{q}) \approx 3$. Then using \eqref{eq:orderparameter}, we find that
\beq
	H_{{\rm h},{\rm int}} \approx -\frac{3 V}{2} \sum_{\mathbf{r}_i} \Phi(\mathbf{r}_i)^2.
\eeq

We proceed by decoupling this term with a scalar field by a Hubbard-Stratonovich transformation, using the identity
\beq
	\exp\left(\frac{3 V}{2} \Phi^2\right) \propto \int \mathcal{D} \phi \, \exp\left( - \frac{1}{6} \phi^2 - \sqrt{V} \phi \Phi \right)
\eeq
for every $\mathbf{r}_i$ after expressing our theory as a path integral. This implies that we may replace the interaction Hamiltonian by
\beq
	H_{\mathrm{int}} = \sum_{\mathbf{r}_i} \left(\frac{1}{6} \phi(\mathbf{r}_i)^2 + \sqrt{V} \Phi(\mathbf{r}_i) \phi(\mathbf{r}_i) \right).
\eeq
We note that $\phi$ transforms in the same way as the order parameter $\Phi$, and it will act as the order parameter in what follows. We also expect that $\phi$ has a uniform condensate in the ordered phase, so in the scaling limit we only need to consider its Fourier components close to zero momentum. Concentrating on the interaction term,
\bea
	\sum_{\mathbf{r}_i}\Phi(\mathbf{r}_i) \phi(\mathbf{r}_i) &=& \int \frac{\dd^2 k \, \dd^2 k'}{\left( \mathcal{A}_{\mathrm{BZ}} \right)^2} \phi(\mathbf{k})	\Big[ c^{\dagger}_{\mathrm{k}' + \mathrm{k},A} c^{\phantom{\dagger}}_{\mathrm{k}',A} \E^{\I \mathbf{k} \cdot \mathbf{s}_1/2} \nn
	&& \qquad - \ c^{\dagger}_{\mathrm{k}' + \mathrm{k},B} c^{\phantom{\dagger}}_{\mathrm{k}',B} \E^{-i \mathbf{k} \cdot \mathbf{s}_1/2} \Big],
\eea
the only regions of interest at low energy are $\mathbf{k} \sim 0$ and $\mathbf{k}' \sim \mathbf{K}, \mathbf{K}'$. Using Eqs.~(\ref{eq:spinordef})-(\ref{eq:gamma_rep}) and going back to real space, we find
\beq
	H_{\mathrm{int}} = \sqrt{V'} \int \dd^2 x \, \phi(\mathbf{x}) \overline{\Psi}(\mathbf{x}) \Psi(\mathbf{x}) + \frac{1}{6} \int \dd^2 x \phi(\mathbf{x})^2,
\eeq
where positive numerical constants have been absorbed into the definition of $V' \propto V$.

To summarize, our full Hamiltonian is
\bea
	H_{\rm h} &=& \int \dd^2 x \, \overline{\Psi}(\mathbf{x}) \left( v_\text{F} \gamma^x \partial_x + v_\text{F} \gamma^y \partial_y + \sqrt{V'} \phi(\mathbf{x}) \right) \Psi(\mathbf{x}) \nn
	&& + \ \frac{1}{6} \int \dd^2 x \, \phi(\mathbf{x})^2\,.
\eea
A renormalization group transformation will generate dynamical terms for the scalar field as well as all interactions allowed by symmetry, resulting in the GNY QFT of \eqref{eq:LGNY} modulo irrelevant terms. Alternatively, one may integrate out the $\phi$ field entirely at this stage and obtain the purely fermionic GN model of \eqref{eq:LGN}. This establishes that the phase transition of $H_{\rm h}$ lies in the chiral Ising universality class, and gives an explicit method for relating its symmetries to those in field theory.

\section{Spectrum in $1/N$ and $D-2$ Expansions}
\label{app:largen}

In this Appendix, we detail the spectrum of the chiral Ising universality class using the $1/N$ and $\epsilon' = d-1$ expansions. A major benefit of the $1/N$ expansion is that one may easily obtain every state in the spectrum for both $g=g_c$ and for relevant deviations $g \neq g_c$ exactly in the $N=\infty$ limit. This allows us to compute crossovers from the critical point to the proximate phases. We will also show that the $N\rightarrow \infty$ limit of the torus spectrum commutes with the $\epsilon \rightarrow 0$ and $\epsilon' \rightarrow 0$ limits to leading order.

\subsection{$1/N$-expansion}

We will find it convenient to work with the GN model, \eqref{eq:LGN}. After decoupling the interaction by a Hubbard-Stratonovich transformation, the imaginary-time Lagrangian is given by
\begin{equation}
    \mathcal{L}_\text{GN} = - \overline{\Psi}^j \left( \cancel\partial + \phi \right) \Psi^j + \frac{N }{2 \tilde{g}} \phi^2,
\end{equation}
where we have defined $g = \tilde{g}/N$, expecting that $\tilde{g}$ remains finite for $N \rightarrow \infty$. As in the main text, we have $N_f$ flavors of $n_\text{D}$-component Dirac fermions, and define $N = n_\text{D} N_f$ to be the total number of degrees of freedom. We will take $n_\text{D}$ to be finite, taking $N_f$ (and therefore $N$) to infinity.

We may now integrate out the fermions exactly, obtaining the Euclidean action
\beq
	\mathcal{S}_{\phi} = - N_f \, \mathrm{Tr} \ln \left( \cancel\partial + \phi  \right) + \frac{N}{2 \tilde{g}} \int \dd^{d+1} x \, \phi^2.
	\label{eq:sphi}
\eeq
For $N \rightarrow \infty$, the path integral may be evaluated at its saddle-point. Defining $\Delta = \langle \phi \rangle$ in this limit, the saddle-point configuration is given by the gap equation
\beq
	\frac{1}{\mathcal{A}} \sum_k \int \frac{\dd \omega}{2 \pi} \frac{1}{\omega^2 + k^2 + \Delta^2} = \frac{1}{\tilde{g}},
\eeq
which is to be solved for $\Delta$.  Both sides of this equation are divergent and regularization-dependent, but if we consider deviations from the critical coupling, we obtain
\beq
	\int \frac{\dd p}{2 (2 \pi)^d}p^{d-2} - \frac{1}{2\mathcal{A}} \sum_k \frac{1}{\sqrt{ k^2 + \Delta^2}} = \frac{1}{\tilde{g}_c} - \frac{1}{\tilde{g}}.
	\label{eq:gapeqn}
\eeq
The left-hand side of \eqref{eq:gapeqn} is finite for $1<d<3$. From this equation, we may solve for the energy gap $\Delta$ by writing
\beq
	\sqrt{\tau_2} \, g_{1/2}^{(d)}(\Delta,\tau) = - 2 \pi \sqrt{\mathcal{A}}(\tilde{g}_c^{-1} - \tilde{g}^{-1}),
	\label{eq:gapeqn2}
\eeq
where we define the special function
\bea
	g_{s}^{(d)} &=& \frac{\pi^s}{\Gamma(s)} \Bigg\{ \int_1^{\infty} \dd \lambda \, \lambda^{s-1} \exp\left( - \frac{\lambda \tau_2 \mathcal{A} \Delta^2}{4 \pi} \right) \Theta\left( \lambda, \mathbf{\Omega}(\tau)  \right)^{d/2} \nn
	&& + \ \tau_2^{-d/2} \int_1^{\infty} \dd \lambda \, \lambda^{d/2 - s - 1} \Big[ \Theta\left( \lambda, \mathbf{\Omega}(\tau)^{-1}  \right)^{d/2} - 1 \nn
	&& - \ \frac{\tau_2 \mathcal{A} \Delta^2}{4 \pi \lambda} \Big] + \frac{\tau_2^{-d/2}}{s - d/2} - \frac{\mathcal{A} \Delta^2}{4 \pi} \frac{\tau_2^{1-d/2}}{1 + s - d/2} \Bigg\} ,
	\label{eq:g12}
\eea
and the Riemann Theta function $\Theta\left( \lambda, \mathbf{\Omega}(\tau)  \right)$ was defined in Eqns.~(\ref{eq:rthet1})-(\ref{eq:rthet2}). We note that $\Delta$ is a monotonically increasing function of $g$, and that at $g = g_c$, it only depends on the shape of the torus. This gap equation is identical to that obtained in the large-$N$ limit of the Wilson-Fisher CFT~\cite{Whitsitt2016}.

We may now write the saddle-point Lagrangian as
\beq
	\mathcal{L}_{\pm} = - \overline{\Psi}^j \left( \cancel\partial \pm \Delta \right) \Psi^j + \frac{N }{2 \tilde{g}} \Delta^2.
\eeq
Because the gap equation only depends on $\Delta^2$, we have a sign ambiguity in choosing $\Delta = \langle \phi \rangle$. Unlike the Wilson-Fisher case~\cite{Whitsitt2016}, this sign difference is physical, and it has an effect on the torus spectrum. In particular, the full set of states are those obtained with positive $\Delta$ \emph{and} those obtained for negative $\Delta$. But this sign has no effect on the spectrum of the theory, so we simply have two copies of a free massive Dirac spectrum with gap $|\Delta|$. Both sectors of the theory have the same ground-state energy, which may be calculated by temporarily taking a finite length $0<\tau<T$ in the Euclidean-time partition function $\mathcal{Z}_{\pm}(T) = \int \mathcal{D}\overline{\psi}\mathcal{D}\psi \exp(-\int_0^T d\tau \int d^d x \mathcal{L}_{\pm})$, and then calculating $E_0 = -\lim_{T \rightarrow \infty} T^{-1} \ln \mathcal{Z}_{\pm}(T)$. This results in
\bea
	E_0 &=& - \frac{N}{2} \sum_k \sqrt{k^2 + \Delta^2} + \frac{N \mathcal{A}  (\tilde{g}_c^{-1} - \tilde{g}^{-1})}{2} \Delta^2 \nn
	&=& \frac{2 \pi}{\sqrt{\tau_2} \mathcal{A}} g^{(d)}_{-1/2}(\Delta,\tau) + \frac{N \mathcal{A}}{2\tilde{g}} \Delta^2,
\eea
where in the second line we have used $g_c^{-1} = 0$ in dimensional regularization, and the special function is defined in \eqref{eq:g12}. We note that the ground state energy is actually ambiguous up to an overall constant, but our choice is such that the energy density vanishes at $g = g_c$ and $\mathcal{A} = \infty$.

To summarize, most of the states in the spectrum are described by the Hamiltonian
\beq
	H_{N} = E_0 + \sum_{k,s,j} \sqrt{k^2 + \Delta^2} \left( b_s^{ \, j\dagger}(k) b^j_s(k) + c_s^{ \, j \dagger}(k) c^j_s(k) \right),
	\label{eq:largen_spec}
\eeq
where $s = 1,...,n_\text{D}/2$ and $j = 1,...,N_f$. In addition to the degeneracies for different values of $k$, $s$, and $j$, every state is doubled due to the different values of $\pm \langle \phi \rangle$. The ground state has energy $E_0$ and is two-fold degenerate, while the first excited state has energy $E_0 + \Delta$ and degeneracy $2N$. We note that this Hamiltonian has an emergent SU($N$) symmetry, where the $N$ operators $b^j_s(k)$ and $c^j_s(k)$ together transform in the fundamental representation.

However, the Hamiltonian $H_N$ does not describe every state in the spectrum. The excited states which are singlets under the SU($N$) symmetry are instead described by fluctuations of the scalar mode $\phi$. We may obtain their energy by writing $\phi = \Delta + \tilde{\phi}/\sqrt{N}$ and expanding \eqref{eq:sphi}. To leading order in $1/N$, we obtain (ignoring a constant)
\beq
	\mathcal{S}_{\phi} = \frac{1}{2} \int \frac{\dd \omega \, d^d k}{(2 \pi)^d} \Pi(\omega,k) |\tilde{\phi}(\omega,k)|^2,
\eeq
where
\begin{widetext}
	\beq
	\Pi(\omega,k) = \int \frac{\dd \Omega}{2 \pi} \sum_q \frac{2 \Delta^2 + \omega^2 + k^2 + \omega \Omega + k \cdot q}{\left( \Omega^2 + q^2 + \Delta^2 	\right) \left[ (\omega + \Omega)^2 +  (q + k)^2 + \Delta^2 \right]}.
	\label{eq:Pi_def}
\eeq
\end{widetext}
The function $\Pi(\omega,k)$ is the inverse Euclidean propagator of the $\tilde{\phi}$ field, so solving $\Pi(i \omega = E_{\phi}(k),k) = 0$ for $E_{\phi}(k)$ gives the energy splitting of the states created by the scalar field from the ground state. The integrals and sums are convergent, and one can solve this equation numerically~\cite{Whitsitt2017,Thomson2017}. Each solution obtained this way has a two-fold degeneracy corresponding to the sign choice $\pm \langle \phi \rangle$. Together with \eqref{eq:largen_spec}, this describes the complete spectrum at $N=\infty$.

\subsection{$\epsilon'$-expansion}

We now consider the expansion in $d = 1 + \epsilon'$ spatial dimensions, which we only study at leading order. The critical Hamiltonian associated with \eqref{eq:LGN} is
\beq
	H_\text{GN} = \int \dd^d x \left[ \overline{\Psi}^j  \cancel{\nabla} \Psi^j - \frac{g^{\ast}}{2} \left( \overline{\Psi} \Psi \right)^2 \right],
\eeq
where $g^{\ast} = 2 \pi \epsilon'/(N-2)$ to leading order in $\epsilon'$~\cite{zinn1991}. The computation proceeds as in Section \ref{sec:eps_expansion}, but is simpler due to the lack of bosonic modes. We find the zero-mode effective Hamiltonian
\beq
	h_{\mathrm{eff},k=0} = - \frac{\pi \epsilon'}{\sqrt{\mathcal{A}}(N-2)} Q^2,
	\label{eq:eps'_effham}
\eeq
where $Q$ has the same definition as in Section \ref{sec:eps_expansion}, leading to the same integer eigenvalues $Q \in [-N/2,N/2]$ and degeneracy \eqref{eq:q_deg}. The Hamiltonian also has a $Q \rightarrow -Q$ symmetry, so the degeneracy analysis of the lowest two states is identical to that of the $\epsilon$ expansion. Extrapolating to $\epsilon' = 1$ for the $N=4$ case results in the predictions $\sqrt{\mathcal{A}} E_1 = 3 \pi/2 \approx 4.7$ and $\sqrt{\mathcal{A}} E_2 = 2 \pi \approx 6.3$ for the energies of the first and second excited states respectively. States with higher energy have a spectrum given by effective Hamiltonians around higher Fock states.

The above expressions explicitly fail for the $N=2$ case. For $D=N=2$, our model is commonly referred to as the massless Thirring (Luttinger) model in the high energy (condensed matter) literature. This case is special as it does not have a phase transition; there is instead a single gapless Luttinger liquid phase \cite{zinnbook}. It is also known that the $N=4$, $D=2$ model exhibits a Kosterlitz-Thouless transition at $g = 0$ \cite{Witten1978}, which is not captured by the $\epsilon'$ expansion. Therefore, as in the case for the O($N$) model, the expansion close to $D=2$ and small values of $N$ likely requires the consideration of non-perturbative effects to obtain the correct critical behavior, which would affect the accuracy of the $\epsilon'$ expansion for smaller values of $N$ \cite{Cardy1980}.

\subsection{Correspondence between expansions}

The large-$N$ expansion for the torus spectrum is much simpler than the $\epsilon$ or $\epsilon'$ expansions in that the zero-momentum modes do not play a special role, so perturbation theory for the spectrum takes a similar form as perturbation theory for the critical exponents. Here, we show that all three expansions are compatible with each other, at least at leading order for low-lying states. We note that the large-$N$ gap equation, \eqref{eq:gapeqn2}, may be solved in any dimension $d$, and in particular it may be solved analytically at leading order in either $\epsilon$ or $\epsilon'$. An explicit computation finds the energy gap at criticality to be
\bea
	\sqrt{\mathcal{A}}\Delta &=& \left( 4 \pi^2 \epsilon \right)^{1/3}, \quad (N \rightarrow \infty, \epsilon \rightarrow 0), \nn
	\sqrt{\mathcal{A}}\Delta &=& \pi \epsilon', \qquad \qquad (N \rightarrow \infty, \epsilon' \rightarrow 0).
	\label{eq:largen_gaps}
\eea
The rest of the spectrum is then given by \eqref{eq:largen_spec} (with the exception of the singlet states, which we do not obtain analytically).

We now consider the large $N$ limit of the $\epsilon$ expansion derived in Section \ref{sec:eps_expansion}. As argued there, the energy spectrum on the torus is given by solving the Hamiltonians $h_{k=0}^{(Q)}$ in \eqref{eq:effham_red} for $Q = 0, 1, ..., N/2$, with the lowest energies given by values of $Q$ close to $N/2$. Therefore, to take the large-$N$ limit, it makes sense to take 
\beq
	Q = \frac{N}{2} - q, \qquad q \ll N,
	\label{eq:def_q}
\eeq
to concentrate on the low-energy spectrum.

At order $\epsilon^{1/3}$, the Hamiltonians $h_{k=0}^{(Q)}$ are shifted harmonic oscillators with a minimum at $\varphi = -(6 \sqrt{Y} Q/U)^{1/3}$, which grows as $\sqrt{N}$ at large $N$. If we expand the Hamiltonians around this minimum with the assumption in \eqref{eq:def_q}, we find that they reduce to simple harmonic oscillators:
\bea
	h_{k=0}^{(q)} &=& \frac{\epsilon^{1/3}}{\sqrt{\mathcal{A}}} \Bigg[ - \frac{3 N \pi^{2/3}}{2^{7/3}} - 3 \left( 4 \pi^2 \right)^{1/3} + q \left( 4 \pi^2 \right)^{1/3} \nn
	&& - \ \frac{1}{2} \frac{d^2}{d \varphi^2} + 6 \left( 2 \pi^4 \right)^{1/3} \varphi^2 + \cdots \Bigg].
\eea
Here, we have used the large-$N$ expansions of the couplings $U$ and $Y$. The anharmonic terms in the effective Hamiltonians vanish at $N=\infty$, so we may simply read off the low-energy spectrum from the well-known oscillator spectrum. The $q$-independent terms in the first line contribute to the universal part of the ground state energy, which is negative as expected for bosonic fields. With respect to the ground state, we see that we have two towers of free ``particle-like'' excitations, with energies given by
\bea
	\sqrt{\mathcal{A}} \Delta_1 &=& \left( 4 \pi^2 \epsilon \right)^{1/3}, \nn
	\sqrt{\mathcal{A}} \Delta_2 &=& \left( 3 \sqrt{6} \pi^2 \epsilon \right)^{1/3}.
\eea
The first excitation agrees with the large-$N$ gap given by solving the gap equation, \eqref{eq:largen_gaps}. For the spectrum to match the large-$N$ results, this value of $\Delta_2$ must therefore be equal to the smallest solution of the equation $\Pi( i \omega = \Delta_2,k=0 ) = 0$, with $\Pi$ defined in \eqref{eq:Pi_def}.

The large-$N$ limit of the $\epsilon'$ expansion is especially simple: we just take the limit \eqref{eq:def_q} in \eqref{eq:eps'_effham}. For $N \rightarrow \infty$, this gives a set of effective Hamiltonians
\beq
	h_{\mathrm{eff},k=0} = - \frac{\pi}{4 \sqrt{\mathcal{A}}} (N+2) \epsilon' + \frac{\pi q}{\sqrt{\mathcal{A}}} \epsilon'.
\eeq
This results in a tower of Fock states with mass $\sqrt{\mathcal{A}} \Delta = \pi \epsilon'$, in agreement with \eqref{eq:largen_gaps}. Since the $\epsilon'$ expansion uses a purely fermionic theory, the bosonic singlet states will necessarily come from an effective Hamiltonian about a higher Fock state with an even number of fermions.


\bibliography{library_gny}

\end{document}